\documentclass[11pt,letterpaper]{article}
\pdfoutput=1
\usepackage{jcappub}
\usepackage{color}
\usepackage{graphicx}
\usepackage{wrapfig}

\usepackage{cancel}

\usepackage{verbatim}
\usepackage{amsmath}
\usepackage{amssymb}
\usepackage{gensymb}
\usepackage{subfig}
\usepackage{url}
\usepackage{bbold}
\usepackage{xspace}

\usepackage{multirow}
\usepackage{threeparttable}
\usepackage{paralist}

\usepackage{color}
\definecolor{darkgreen}{rgb}{0,0.5,0}

\newcommand{\GeV}{\text{GeV}}

\newcommand{\be}{\begin{equation}}
\newcommand{\ee}{\end{equation}}

\usepackage{accents}
\newlength{\dhatheight}

\newcommand{\gtap}{\gtrsim}
\newcommand{\ltap}{\lesssim}
\newcommand{\eg}{{\it e.g.}}
\newcommand{\ie}{{\it i.e.}}

\usepackage{color}
\begin{document}
\begin{flushright}
FERMILAB-PUB-14-411-T\\
CERN-PH-TH-2014-219
\end{flushright}
\title{WIMPs at the Galactic Center
}

\author[a]{Prateek  Agrawal,}

\affiliation[a]{Theoretical Physics Department, Fermilab, Batavia, Illinois 60510, USA}

\author[b]{Brian Batell,}

\affiliation[b]{CERN, Theory Division, CH-1211 Geneva 23, Switzerland}
 
\author[a]{Patrick J.  Fox}

\author[a]{and Roni  Harnik}

\emailAdd{prateek@fnal.gov}
\emailAdd{brian.batell@cern.ch}
\emailAdd{pjfox@fnal.gov}
\emailAdd{roni@fnal.gov}

\date{\today}


\arxivnumber{}

\abstract{
Simple models of weakly interacting massive particles (WIMPs) predict dark matter annihilations into pairs of electroweak gauge bosons, Higgses or tops, which through their subsequent cascade decays produce a spectrum of gamma rays. Intriguingly, an excess in gamma rays coming from near the Galactic center has been consistently observed in Fermi data.  A recent analysis by the Fermi collaboration confirms these earlier results. Taking into account the systematic uncertainties in the modelling of the gamma ray backgrounds, we show for the first time that this excess can be well fit by these final states.  In particular, for annihilations to ($WW$, $ZZ$, $hh$, $t\bar{t}$), dark matter with mass between threshold and approximately (165, 190, 280, 310) GeV gives an acceptable fit.  The fit range for $b\bar{b}$ is also enlarged to $35\,\GeV\ltap m_\chi\ltap 165$ GeV.  These are to be compared to previous fits that concluded only much lighter dark matter annihilating into $b$, $\tau$, and light quark final states could describe the excess.  We demonstrate that simple, well-motivated models of WIMP dark matter including a thermal-relic neutralino of the MSSM, Higgs portal models, as well as other simplified models can explain the excess.
}

\maketitle

%%%%%%%%%%%%%%%%%%%%%%%%%%%%%%

\newpage
\section{Introduction}

The weakly interacting massive particle (WIMP) is a well-motivated
scenario for dark matter.  It explains the observed abundance of dark
matter as arising from thermal freeze out of the WIMP --- a particle with a mass
and annihilation cross section at the weak scale.  Taken literally,
this picture suggests that dark matter (DM) particles participate in weak
interactions and/or the mechanism for electroweak symmetry breaking.
This motivates the annihilation of dark matter to  $W$ and $Z$ bosons, and
to the particle responsible for electroweak symmetry breaking, the Higgs boson. In models motivated by the hierarchy problem dark matter interactions could be mediated by top-partners which would lead to annihilation into tops.
In this work, we
study the consequence of dark matter annihilations to electroweak bosons
and top quarks.

\begin{figure}[t] %  figure placement: here, top, bottom, or page
   \centering
   \includegraphics[width=0.485\columnwidth]{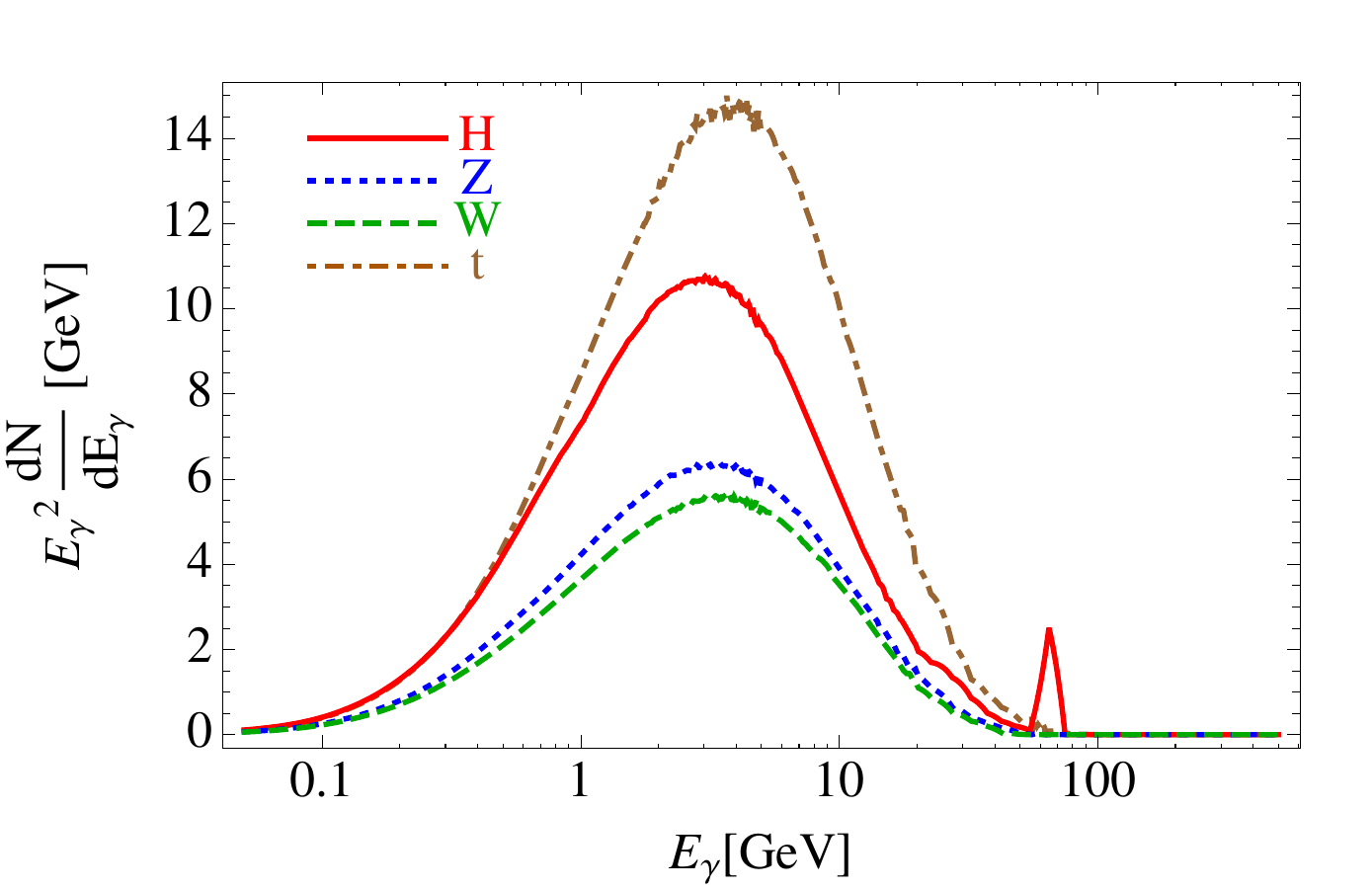} 
   \includegraphics[width=0.485\textwidth]{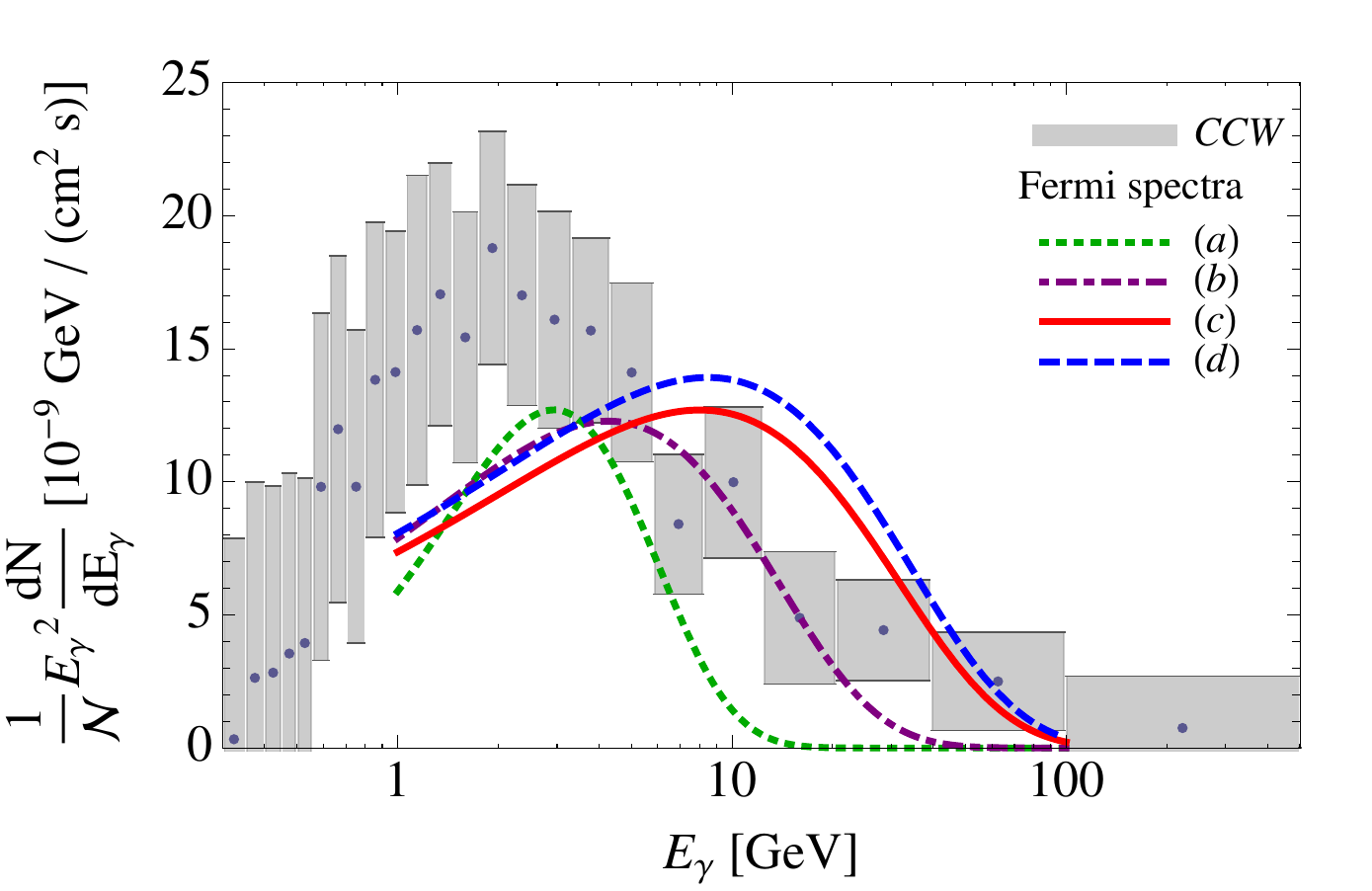} 
   \caption{\emph{Left:} The $\gamma$-ray spectrum produced by a single $W$, $Z$, Higgs boson, and top quark, decaying  at rest, weighted by $E_\gamma^2$. 
   \emph{Right:}    The residual spectrum of the Galactic center excess 
    taken from
   \cite{Calore:2014xka}. The error bars only show the diagonal part
   of the covariance matrix, and have a large degree of
   correlation between them. We also show the four best-fit spectra
from the Fermi analysis~\cite{simonatalk} which fit the excess well. We will dub these, from softest to hardest, as Fermi spectra (a) through (d). The normalization $\mathcal{N}$ corrects for the difference in the region of interest between
the two analyses.
   }
   \label{fig:at-rest}
\end{figure}

Gamma rays are a promising signal of a $W$, $Z$, Higgs or top
produced in the sky. The decay products of these particles yield
photons through final state
radiation and neutral pion decays, which travel unimpeded to our
telescopes.
In the left panel of Figure~\ref{fig:at-rest} we show the
differential gamma ray flux from $W$, $Z$, $h$ and top decaying at rest,
multiplied by the photon energy squared. The spectra are generated by
{\tt PYTHIA}~\cite{Sjostrand:2006za}.  The multiplication by $E^2$ is
traditional for astrophysical fluxes and accounts for the steeply
falling spectra of astrophysical backgrounds. All three fluxes are
qualitatively similar, with a peak in the vicinity of 2--5 GeV.
A notable difference is the sharp peak, at $m_h/2$, coming from the decay
$h\to\gamma\gamma$. Though this decay is rare, the photons it produces
are hard and this spectral feature is enhanced by the factor of $E^2$.
This feature, however, is highly sensitive to small boosts of the Higgs
and to the energy resolution of the detector. 

Interestingly, repeated analysis of Fermi data,
starting from Goodenough and
Hooper~\cite{Goodenough:2009gk,Hooper:2010mq} and
by many subsequent
groups~\cite{Boyarsky:2010dr,Hooper:2011ti,Linden:2012iv,Abazajian:2012pn,Hooper:2013rwa,Gordon:2013vta,Abazajian:2014fta,Daylan:2014rsa,Zhou:2014lva,Calore:2014xka},
has uncovered an excess in gamma rays from the center of the Milky Way,
peaking around 2--3 GeV.  This is known as the Galactic Center Excess (GCE), or the Gooperon.
Though the excess is highly statistically significant, it relies on
a diffuse background model developed by the
Fermi collaboration. This background model provides a reasonble fit to
the diffuse emission over the whole sky, but it is not clear that it is
appropriate for the Galactic center. 
 This systematic uncertainty on the
background dominates the statistical one and is difficult to
quantify. 
A first attempt at accounting for systematics was made
by Calore, Cholis and Weniger (CCW)~\cite{Calore:2014xka} through the study of a large number of Galactic diffuse emission models and analysis of the residuals in test regions %by measuring correlations 
along the signal
poor regions of the Galactic disk.

Very recently, the Fermi collaboration has performed an analysis of a
region around the Galactic center considering several different
dedicated background models~\cite{simonatalk}. 
For each background model, the fits improve significantly when a
spherically symmetric component -- with a morphology and spectrum
similar to that of annihilating dark matter -- is added.
In the right panel of Figure~\ref{fig:at-rest} we show the
DM-like spectra found by Fermi assuming four different
background models. Their envelope is a good representation of the systematic
uncertainties, but does not account for the
fit uncertainties. In the same plot we also show the
residual spectrum found by CCW which
includes both statistical as well as some systematic effects. We see
that the two spectra are consistent with one another once we account
for the difference in their region of interest. They are also broadly
consistent with the 2--3 GeV excess reported in earlier analyses.
However, in a continuing trend,
inclusion of systematic uncertainties has allowed for an increasingly 
harder spectrum in the new analyses. It is thus important to consider
a broad array of motivated dark matter models which can explain the 
excess once full uncertainty on the spectrum is taken into account.

It is interesting to notice the rough similarity between the two
panels of Figure~\ref{fig:at-rest}. This observation leads to the
consideration of dark matter models that could explain
the GCE with dark matter annihilating to electroweak bosons or to tops.
In most previous dark matter interpretations of the GCE, starting
with~\cite{Goodenough:2009gk,Hooper:2010mq}, the dark matter was
assumed to annihilate into bottom quarks or $\tau$ leptons. Assuming
these annihilation channels (and without including the new Fermi
uncertainties), the mass of dark matter that best fits the excess is
in the region of 30 to~50~GeV for $b$'s and around 10 GeV for $\tau$
leptons. In addition, dark matter annihilation into new particles
which decay further to $b$'s or jets have been considered. All of of
these options present
interesting model building challenges and several interesting attempts
have been made~
\cite{Logan:2010nw,
Buckley:2010ve,
Zhu:2011dz,
Marshall:2011mm,
Boucenna:2011hy,
Buckley:2011mm,
Hooper:2012cw,
Buckley:2013sca,
Hagiwara:2013qya,
Okada:2013bna,
Boehm:2014hva,
Lacroix:2014eea,
Alves:2014yha,
Berlin:2014tja,
Agrawal:2014una,
Izaguirre:2014vva,
Ipek:2014gua,
Cerdeno:2014cda,
Boehm:2014bia,
Ko:2014gha,
Abdullah:2014lla,
Ghosh:2014pwa,
Martin:2014sxa,
Basak:2014sza,
Berlin:2014pya,
Agrawal:2014aoa,
Cline:2014dwa,
Detmold:2014qqa,
Chang:2014lxa,
Cheung:2014lqa,
Huang:2014cla,
Balazs:2014jla,
Ko:2014loa,
Okada:2014usa,
Ghorbani:2014qpa,
Bell:2014xta,
Borah:2014ska,
Cahill-Rowley:2014ora,
Yu:2014pra,
Cao:2014efa,
Freytsis:2014sua,
Heikinheimo:2014xza}, 
mostly for annihilation to $b$'s, $\tau's$ and jets.

We find that WIMP dark matter annihilating to $W$'s, $Z$'s,
Higgses, or tops,
can fit the observed excess reasonably well. We show that this is the
case for the spectra found in~\cite{Calore:2014xka}, and this result
is reinforced by the recent Fermi result.  In particular, if we take
the union of the preferred regions for each analysis, we find
that the range of DM masses can extend well
above what was previously thought. We show a summary of the results in
Table~\ref{tab:massrange}.
This opens up several simple dark matter model building avenues for
the GCE. It was noted that the simplest supersymmetric models with a
thermal relic fail to fit the signal~\cite{Cahill-Rowley:2014ora}
assuming annihilation into bottom quarks.
We will find that once electroweak gauge bosons are considered, the
signal may be explained within the MSSM.

\begin{table}
\begin{center}
\begin{tabular}{|c|c|c|}
\hline
channel & from & to [GeV]\\
\hline
$WW$ &  $m_W$   &  165  \cr
$ZZ$ &   $m_Z$  &  190    \cr
$hh$ & $m_h$    & 280     \cr
$t\bar t $ &  $m_t$   &  310    \cr
$b\bar b$ &  35 GeV   &  165    \cr
\hline
\end{tabular}
\end{center}
\caption{The allowed dark matter mass range for $\chi \chi \to X X$, with $X=\{h, W^\pm, Z, t, b\}$, found by combining the preferred regions from the results of our fits to the GCE.}\label{tab:massrange}
\end{table}

We begin by reviewing features of the photon flux from dark matter
annihilation in Section~\ref{sec:target}, focusing on relevant
inputs which affect the rate and shape of the flux. 
In
Section~\ref{sec:fits} we describe the excess seen by the CCW
\cite{Calore:2014xka} and Fermi~\cite{simonatalk} analyses, and
present fits to the GC excess in the mass
versus cross section plane for the final states described above.  In Section~\ref{sec:models} we
discuss several simple models which lead to dark matter annihilation
into weak gauge bosons, Higgses or tops.
We conclude in section~\ref{sec:conclusion}.

\section{Dark Matter Annihilation at the Galactic Center}\label{sec:target}

While DM can annihilate directly to a pair of hard photons, this
process is typically loop suppressed.
The production of photons is dominated by production of SM particles
which subsequently produce photons through decays, or to a lesser
extent bremsstrahlung.  The differential flux of such photons from a
given direction $\psi$ is given by,
\be
\frac{d N}{d\Omega dE} (\psi) 
= \frac{1}{4\pi\eta}\frac{f^2_\chi J(\psi)}{m_\chi^2}\sum_i 
\langle \sigma v\rangle_i\frac{dN^i}{dE_\gamma}~, 
\label{eq:flux}
\ee
with $\eta=2(4)$ for %Majorana (Dirac) 
self-conjugate (non-self-conjugate) DM. The quantity
$dN^i/dE_\gamma$ is the spectrum of photons obtained per annihilation
for the final state $i$. The line-of-sight integral,
$J(\psi)$, is given by
\be
J(\psi)=\int_{\mathrm{l.o.s.}} ds\, \rho(r)^2~\,,
\label{eq:flux-J}
\ee
where $r$ is the distance from the Galactic center.  The quantity $f_\chi$ is the fraction of dark matter that is %$\chi$ is 
doing the annihilation. For simplicity we will assume only one species $\chi$ is annihilating, but the formalism can be trivially generalized to many by taking a sum.

In this section we will discuss each of the factors in~\eqref{eq:flux} in turn, paying attention to the uncertainties and their relation to dark matter properties. We will begin with the line-of-sight integral, $J(\psi)$, continue with the annihilation fraction $f_\chi$, and then discuss the spectra $dN^i/dE_\gamma$ in Section~\ref{sec:spectra}.

\subsection{The Line-of-Sight Integral and Halo Uncertainties}\label{sec:halo uncertainties}

Since the dark matter density peaks sharply
towards the center of the Galaxy, the Galactic center is a promising
place to look for dark matter annihilations.
In practice, the backgrounds near the center of the Galaxy are poorly
understood, and it is not possible to perform a model independent
subtraction. One approach that is commonly used is to 
include an additional dark matter component to the fit in addition to
the various background components. It is found that the fit improves
dramatically when such a dark matter component is included.
The photons absorbed by the dark matter template are the residuals,
to which different hypotheses can be compared.

The current sensitivity
to different dark matter profiles is somewhat poor. In order to
compare results across different analyses, it is convenient to define a
canonical dark matter profile. 
A typical choice is the (generalized)
Navarro-Frenk-White (NFW)~\cite{Navarro:1995iw} profile,
\be
\label{eq:NFW}
\rho(r) = \rho_0\frac{\left(r/r_s\right)^{-\gamma}}{\left(1+r/r_s\right)^{3-\gamma}}~.
\ee
As our canonical profile we choose the local dark matter density (at $r=r_\odot \equiv 8.5$ kpc) to be
$\rho_\odot =0.4$ GeV/cm$^3$,  the scale radius $r_s = 20$ kpc and
$\gamma=1.2$.  % for the canonical profile.
This is a convenient parametrization of a
spherically symmetric diffuse component.  The
observational constraints on the actual dark matter halo profile have
relatively large uncertainty, especially near the center of the
Galaxy. This has implications for extracting particle
physics information -- particularly the annihilation rate -- from the signal.

There are a number of measurements of the dark matter density using
different techniques; however, the uncertainty on the dark matter
distribution near the center of the Galaxy remains large.
A complete modeling of the uncertainty in $J(\psi)$ is far
beyond the scope of the paper. We adopt the following method to
estimate this uncertainty. CCW found 
the range of the inner slope for a variety of Galactic Diffuse Emission (GDE) models
to be $\gamma = 1.2\pm0.1$~\cite{Calore:2014xka}. We use this range as the
uncertainty on $\gamma$. The normalization of the profile is
not fixed by the GCE observation. We use the range $\rho_\odot = 0.4\pm0.2$
GeV/cm$^3$ \cite{Iocco:2011jz} to parametrize the normalization. We keep the
values for $r_s=20$ kpc and $r_\odot = 8.5$ kpc fixed.  

The total rate depends on the quantity,
\begin{align}
\bar{J}
&=
\frac{1}{\Delta \Omega}
\int_{\Delta \Omega} J(\psi) d\Omega
\equiv
\mathcal{J}\times
\bar{J}_{\mathrm{canonical}} \,.
\end{align}
Here $\Delta \Omega$ is the region of interest (ROI) for a given analysis, and $\bar{J}_{\mathrm{canonical}}$ is the central value of 
$\bar{J}$.
$\mathcal{J}$ characterizes the profile uncertainty as the deviation from the canonical profile. $\bar{J}_{\mathrm{canonical}}$ and the range of $\mathcal{J}$
depend on the ROI of a particular analysis. The relevant ranges for
analyses we consider will be defined below.

We stress that we consider the
NFW profile as a parametrization of the dark matter profile towards
the Galactic center. In particular, we do not assume in our analysis
that $\rho_\odot$ is the actual local dark matter density.
A proper analysis of the dark matter profile would involve
considering all the observational constraints on the dark matter
together with the Galactic center excess, along the lines of
\cite{Catena:2009mf}. Measurements from Gaia
\cite{Price-Whelan:2013kpa,Bovy:2013} 
would help reduce the uncertainty in a model independent way.

\subsection{The Annihilation Fraction $f_\chi$}

In general, an annihilation signal does not necessarily originate from all of dark matter.
We parametrize this possibility as the fraction, 
\begin{align}
f_\chi
&=
\frac{\Omega_\chi}{\Omega_{DM}}~,
\end{align}
where $\Omega_{DM}\approx0.27$ is the total DM energy density \cite{Ade:2013zuv} and $f_\chi\leq 1$.
The normalization of the excess observed is proportional to $f_\chi^2
\bar{J} \langle \sigma v \rangle$. Direct detection and constraints from neutrino fluxes from the sun are weakened linearly with $f_\chi$, whereas other indirect constraints such as anti-proton fluxes scale quadratically with $f_\chi$.

If the dark matter is produced thermally, then its relic abundance is
directly tied to its annihilation rate in the early universe. Thus,
for a thermal dark matter population we can predict its current 
fraction in the Galaxy from the annihilation rate at freeze out. 
Furthermore, if the annihilation occurs in the $s$-wave (i.e. independent
of initial velocities) and there are no additional (co-)annihilation channels available in the early universe, then the annihilation rate in the early universe
and in the Galactic center today are essentially the same. Thus,
within uncertainties, such a
thermal relic can unambiguously predict the normalization of the
photon spectrum observed.

For a non-thermal population of dark matter, $f_\chi$ is independent of the value of 
$\langle\sigma v \rangle_\mathrm{freezeout}$. Further, there 
 are many known examples where the annihilation rate in the early universe
is different from the rate today. For example, this correspondence fails 
if the annihilation rate is velocity dependent ($p$-wave, Sommerfeld enhanced), or if there are other
annihilation channels available in the early universe
(co-annihilation). These caveats open up a wide class of models which
can be consistent with the observed signal.

\subsection{The Spectrum of Annihilating Dark Matter and the GCE} \label{sec:annihilations}\label{sec:spectra}

Here we discuss the differential energy spectrum produced by dark matter annihilation, $dN/dE_\gamma$.
For the final states we consider, the dominant source of photons in
dark matter annihilation is through the decay of
neutral pions formed in the hadronization and subsequent decay of the annihilation products.
The final states are either color singlets ($W$, $Z$, $h$) or heavy (top) and thus there is no significant amount of QCD radiation in the annihilation interaction itself.
As a result, in order to get the photon energy spectrum we can go to the rest frame of the decay product and boost the distribution back to the rest frame of the annihilation.
Since the dark matter particles themselves are non-relativistic, the
boost of the final state directly correlates with the mass of the dark
matter.

In Figure~\ref{fig:at-rest} we have shown the gamma-ray spectrum
produced by $X$ decaying at rest, where~$X$ is a $W$, $Z$, $h$, or
$t$. The spectra are dominated by neutral pion decays but the small
contribution 
from bremsstrahlung is included as well.
By eye, all of these shapes produce a reasonable fit to the observed
excess. The shapes of these spectra correspond to the case where
the dark matter happens to be very nearly degenerate with $X$. 
In a more generic case the annihilation product will be boosted with a
$\gamma$ factor of $m_\mathrm{DM}/m_X$. The $\gamma$-ray distribution
must thus also be boosted by this amount. 

The direction of a photon emitted in the rest frame of $X$ can be
parametrized by the angle $\theta$ it makes with the velocity of $X$.
Assuming $X$ is not polarized, it will decay isotropically, yielding a
flat distribution in $\cos\theta$. This implies that photons of a
given energy $E_0$ in the rest frame of $X$ lie in a uniform
distribution in the DM rest frame, between energies $E_-$ and $E_+$,
with,
%Assuming $X$ is not polarized, it will decay isotropically. This
%introduces a random angle $\theta$ between every photon and the boost
%it undergoes, which has a flat distribution in $\cos\theta$.
%Taking this angular distribution into account, photons of energy
%$E_0$ in the rest frame of $X$, once boosted, will lie in a flat 
%distribution of photons whose energies are between ~$E_-$ and~$E_+$, with,
\begin{equation}
\label{eq:Epm}
E_\pm=E_0 \gamma(1\pm\beta) = E_0 \left(\frac{m_\mathrm{DM}}{m_X}\pm \sqrt{\frac{m_\mathrm{DM}^2}{m_X^2}-1}\right)\,. 
 \end{equation}
Operationally this means that the distributions in
Figure~\ref{fig:at-rest} (divided by the factor of $E^2$) are
convolved with a unit normalized ``box'' between $E_-$ and $E_+$. If
the photon spectrum in the rest frame of $X$ is $\Phi_X(E)$, in the DM
rest frame the spectrum, $\Phi_\mathrm{DM}(E)$, will be, 
\begin{equation}
\label{eq:boosted}
\Phi_\mathrm{DM}(E) =\frac{1}{2}\int_{-1}^1 dc_\theta\frac{1}{\gamma(1-\beta c_\theta)}\Phi_X\left(\frac{E}{\gamma(1-\beta c_\theta)}\right)= \int_{Ex_-}^{Ex_+} d E' \frac{\Phi_X(E')}{(x_+-x_-)E'}\,,
\end{equation}
where $x_\pm=E_\pm/E_0$, are the smearing factors in equation~(\ref{eq:Epm}) which are independent of photon energy.

\begin{figure}[tp]
  \begin{center}
    \includegraphics[width=0.7\textwidth]{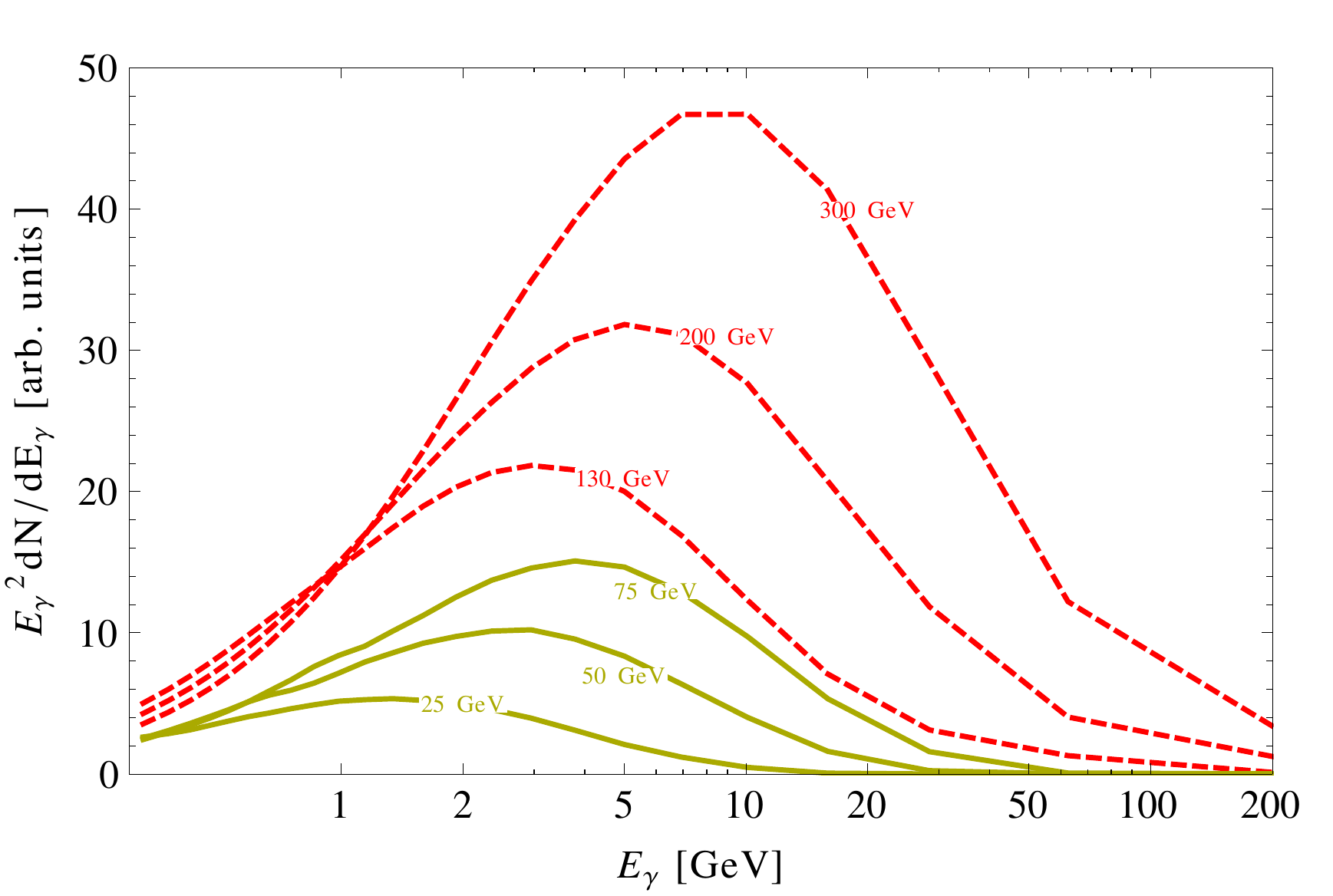}
  \end{center}
  \caption{Spectra of photons arising from $\chi\chi \to hh$ ($\chi\chi \to b\bar{b}$) annihilation shown in dashed red (solid yellow) for a number of different masses of the dark matter.}
  \label{fig:progspec}
\end{figure}

As the dark matter mass grows the boost has the effect of broadening the peak in Figure~\ref{fig:at-rest} and pushing the peak to higher energies. In Figure~\ref{fig:progspec} we show how the $E^2 dN/dE$ spectrum evolves as a
function of the mass of dark matter for the examples of annihilation to
Higgs boson and $b\bar{b}$ final states. Comparing this to the various spectra in the right panel of Figure~\ref{fig:at-rest}, we anticipate that the excess will be well fit over a large region in dark matter mass, particularly once we take into account the range of spectra shown by Fermi. However, even without the new Fermi spectra it is worth noting that dark matter at 130 GeV annihilating to Higgses is quite similar in shape to DM at 50 GeV annihilating to $b\bar b$. Given that the latter is considered a good fit to the GCE, we expect the former to be a good fit as well.

In order to compare annihilations into $W$, $Z$, $h$, or $t$ to previous interpretations we will also show spectra for annihilation into $b\bar b$. Because the $b$ is colored and light, the approximation of going to the rest frame of the $b$-quark and boosting is not valid.  However, the $\gamma$-ray spectrum can be crudely modeled as a log-normal
distribution. The position of the peak in the flux ($E^2
\frac{dN}{dE}$) scales linearly with the mass of dark matter.
Additionally, the value of the flux at the peak also scales linearly with the dark matter mass.
The total fraction of energy carried by the photons is roughly
constant, $\sim \frac14$. To correctly account for hadronization effects, we simulate $b\bar b$ production in {\tt PYTHIA}~\cite{Sjostrand:2006za} with a center of mass energy of $2m_\chi$ to get the spectrum.

\section{Fits to the Galactic Center Excess}\label{sec:fits}

Even though the presence of the excess is relatively robust, in order to make
conclusions about its particle physics origins we need to understand
more detailed properties of the excess. However, the spectral shape
and even the normalization of the excess is somewhat sensitive to the
modeling of the Galaxy, and the dominant uncertainty in the residual
signal is systematic. Correctly estimating the systematic
uncertainties in the background modeling is a challenging task.
The understanding of these backgrounds is
continually evolving, and with that the spectrum of the excess is
changing as well. Therefore, in this analysis we will be conservative in ruling out
models which do not seem to fit the excess perfectly.

As discussed above, the astrophysical constraints on the parameters that determine $J$ are somewhat weak and allow a large range for $\mathcal{J}$, which translates into a large range in the determination of $\langle \sigma v \rangle$ from the observed annihilation rate.
Here we adopt a slightly different presentation from what is usually seen. We define the quantity
\begin{align}
\langle \sigma v \rangle_{\mathrm{eff}} = \mathcal{J} \langle \sigma v\rangle f_\chi^2\, ,
\end{align}
which is directly proportional to the number of annihilations and thus to the observed flux. 
 Fits in the  $m_\chi$--$\langle \sigma v \rangle_\mathrm{eff}$ plane are thus not sensitive to
these uncertainties. Instead, deviations of $\mathcal{J}$ or $f_\chi$ from 1 affect the extracted $\langle \sigma
v\rangle$ values.  Thus, a WIMP with thermal annihilation cross section of
$\langle \sigma v \rangle=2.2\times10^{-26}$ cm${^3}$/s maps on to a
band in $\langle \sigma v \rangle_\mathrm{eff}$, see Figure \ref{fig:chisquaredonshell}.
For ease of comparison with the earlier analyses \cite{Calore:2014xka,simonatalk} we normalise such that $\mathcal{J}=1, f_\chi=1$ corresponds to the case where our $\langle \sigma v\rangle_\mathrm{eff}$ is the equal to the $\langle \sigma v\rangle$ used previously.

We note that we have assumed that the uncertainty on $\mathcal{J}$, particularly the range of allowed slopes $\gamma$ ``factorizes'' from the fit. Namely, we assume that as one varies $\gamma$, the spectral shape of the excess does not change significantly.
It was found~\cite{Calore:2014xka} that within the range of
dark matter profiles we consider this is indeed the case.  We will now consider two fits to the GCE, the first was done by CCW~\cite{Calore:2014xka} and the second is a very recent analysis by the Fermi collaboration~\cite{simonatalk}. These analyses made different assumptions about backgrounds and in the end we consider the union of the preferred regions from these analyses as allowed.
We will first assume that all of the dark matter is made up of the relevant component, which is annihilating, namely $f_\chi=1$. We will return to interpretations of our result for $f_\chi<1$ in section~\ref{sec:wtf}.

\subsection{The CCW Fit}

We use as one benchmark the recent analysis by \cite{Calore:2014xka},
which analyzes
Fermi-LAT data %in the inner $\pm20^{\circ}$ 
in the inner Galaxy over the energy range 300
MeV--500 GeV.  The ROI in \cite{Calore:2014xka} extended to a $\pm20^\circ$ square
around the Galactic center, with the inner $2^\circ$ latitude 
masked out. We find that for the canonical
profile ($r_\odot=8.5$ kpc, $\rho_\odot = 0.4$ GeV/cm$^3$, $r_s =
20$ kpc, $\gamma = 1.20$), the value of $\bar{J}_\mathrm{canonical}= 2.0\times10^{23}$ GeV$^2$ / cm$^5$. Taking into account profile uncertainties discussed in Section \ref{sec:halo uncertainties} the uncertainty in $\bar{J} = \mathcal{J} \bar{J}_\mathrm{canonical}$, yields $\mathcal{J}\in[0.19,3]$.

The analysis of CCW attempts to quantify some of the
systematic uncertainties for the GCE signal, by studying 60 different GDE background models for diffuse emission in the
Galaxy.
They also study the correlations in the spectrum along the signal poor Galactic disk. From this study they extract the residual signal and a set of systematic uncertainties which are correlated across energy bins. 
We show their residual
spectrum in the right panel of Figure \ref{fig:at-rest} obtained for their best fit
GDE model (called ``model F'' in their analysis). The systematic errors
dominate over most of the energy range. Furthermore, the systematic
errors have a large degree of correlation, making it difficult to
visually evaluate the fit quality of the spectra.

This large degree of correlation among the systematic errors is encapsulated by a covariance matrix.  The total error matrix, $\Sigma$, which contains uncorrelated statistical, as well as correlated systematic uncertainties, has been made available by CCW \cite{Calore:2014xka,weiniger-stuff}.  The
systematic uncertainties are well-modeled by expected errors in known
background shapes (for instance $\pi^0 +$ Bremsstrahlung) \cite{Calore:2014xka}.

Using $\Sigma$ it is possible to determine how well a candidate DM model fits the residual spectrum.  To do so we define the goodness-of-fit with a $\chi^2$ test statistic,
\begin{align}
\chi^2
&=
\left[\frac{dN}{dE}(m_\chi,\sigma v) -
\left(\frac{dN}{dE}\right)_{obs}\right]
\cdot \Sigma^{-1} \cdot
\left[\frac{dN}{dE}(m_\chi,\sigma v) -
\left(\frac{dN}{dE}\right)_{obs}\right]\,,
\end{align}
where $m_\chi$ is the mass of the dark matter and $\sigma v$ is its
annihilation rate. 

\begin{figure}[t]
  \begin{center}
    \includegraphics[width=0.8\textwidth]{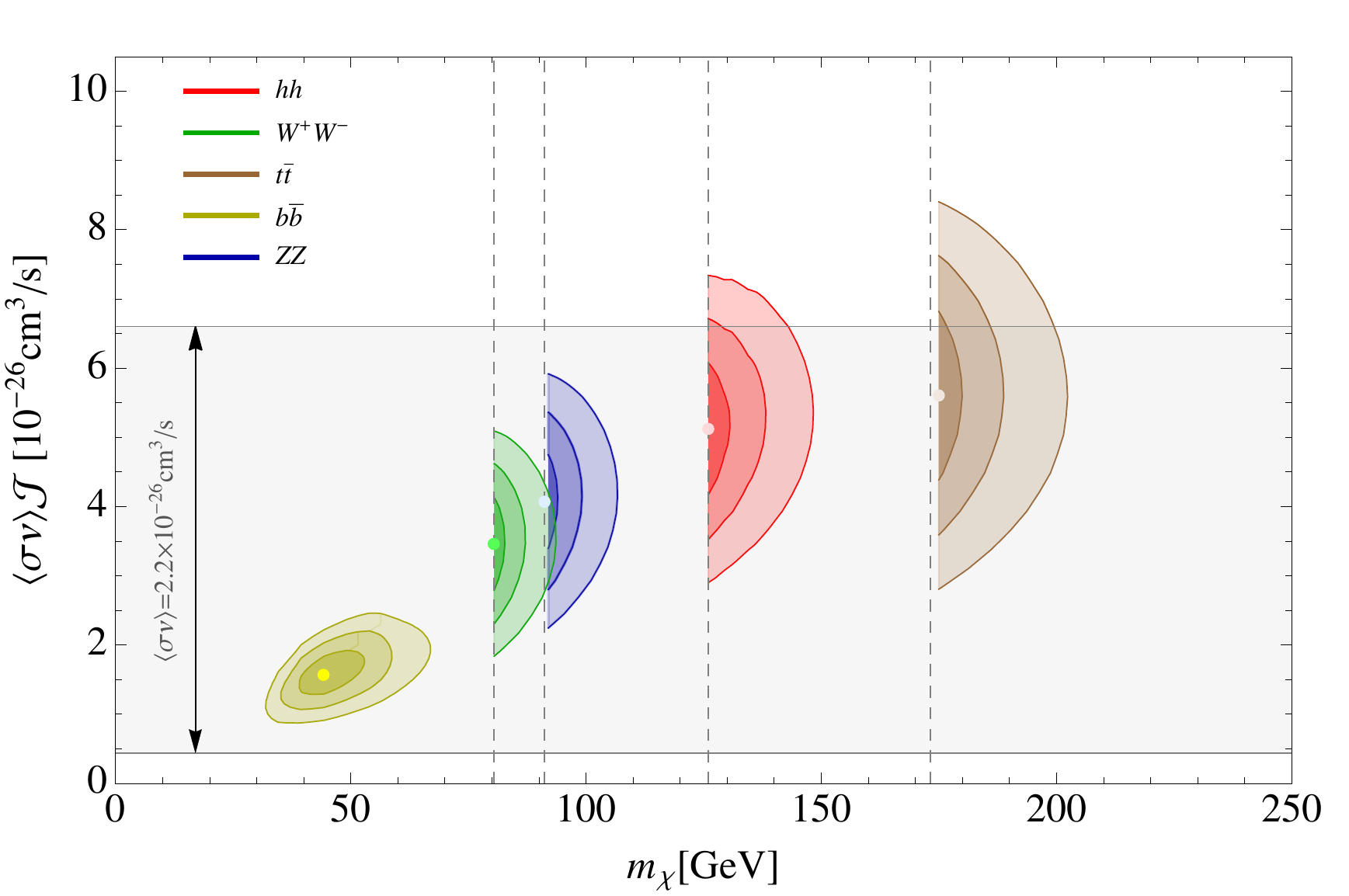}
    \\
    \includegraphics[width=0.7\textwidth]{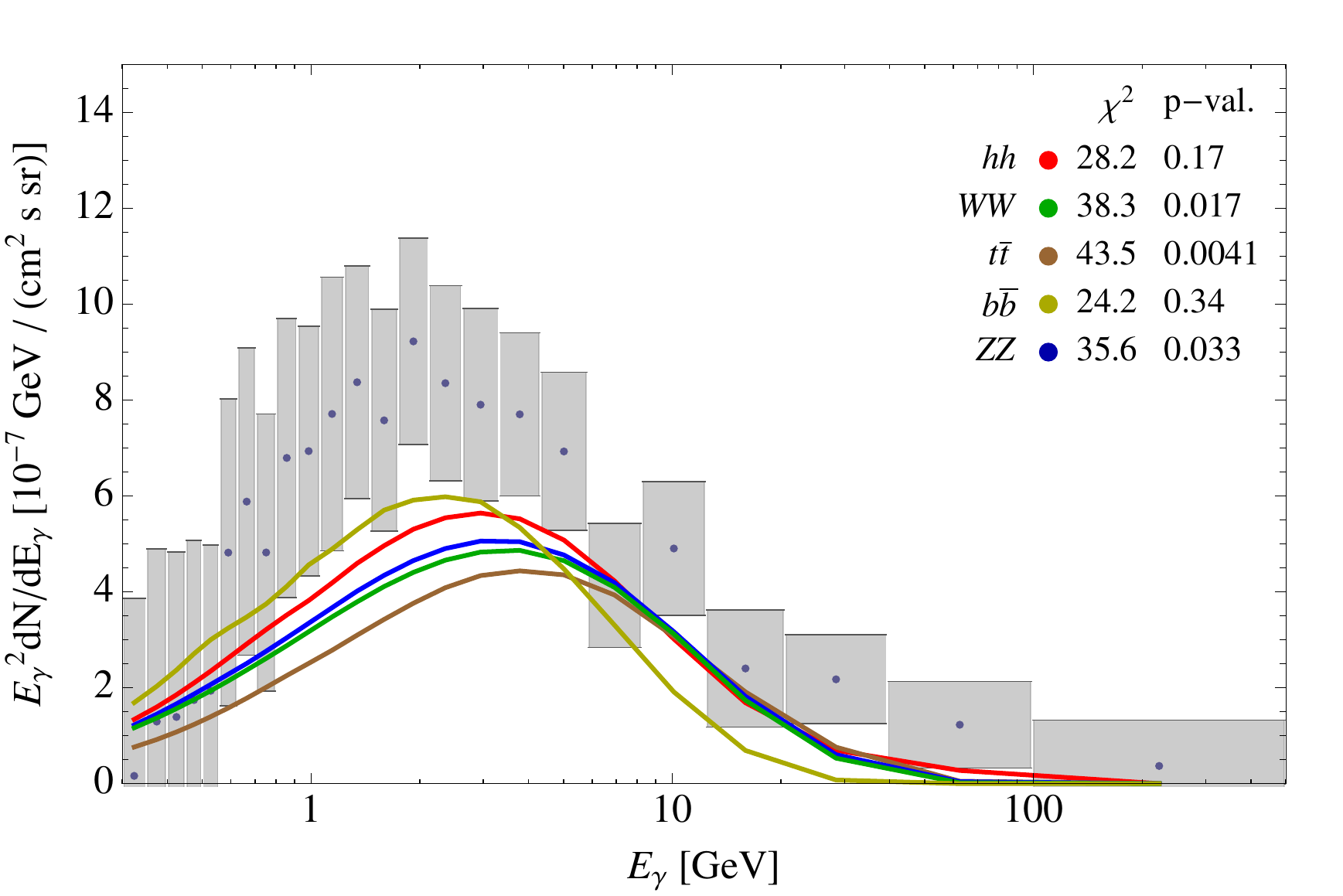}
  \end{center}
  \caption{Top: We show the $\Delta \chi^2$ contours, corresponding to 1,2 and 3$\sigma$, obtained for the hypotheses
  $\chi \chi \to X X$ for $X=\{h, W^\pm, Z, t, b\}$. %The $\Delta\chi^2$ contours correspond to 1,2 and 3$\sigma$ away from the best fit point for that mode. 
  Vertical dashed lines indicate the threshold for each of these final states. The best fit point in each case is indicated.
  Bottom: 
  We show the spectra of photons obtained for the
  corresponding best fit values in the upper plot. The central values and the error bars
  are extracted from
  \cite{Calore:2014xka}. Note that the errors are correlated, and the plotted
  spectra indeed fit the data reasonably well, as indicated by the $\chi^2$ at the best fit.
  }
  \label{fig:chisquaredonshell}
  \label{fig:winningspectra}
\end{figure}

We calculate a number of benchmark spectra for dark matter particles
of varying masses and annihilation cross sections to electroweak
bosons. 
Given the specific final state, we can fit the mass of the dark matter
from the shape information in the GCE. The normalization of the
excess can be 
used to determine the rate of annihilation in
the region of interest. The results of the fit are shown in Figure~\ref{fig:chisquaredonshell}.

In the top panel of Figure
\ref{fig:chisquaredonshell} we show the $\Delta\chi^2$ values for a variety
of SM final states as a function of the dark matter mass and $\langle
\sigma v \rangle_\mathrm{eff}$.  We see that the $b\bar{b}$ and the $hh$
final state spectra have the lowest $\chi^2$ values. The spectra for
other electroweak bosons also fit within $3\sigma$. 
While CCW estimate the systematic uncertainties from the Galactic disk we expect there to be additional, presently unknown, contributions.  This leads us to view the spectra shown as an acceptable fit to the data.
We also show the range of  $\langle \sigma v \rangle_\mathrm{eff}$ which corresponds to a thermal relic cross section.
Within the uncertainty in $\mathcal{J}$ we see that all final states are compatible with a thermal WIMP. 

In the bottom of Figure \ref{fig:winningspectra} we show the best fit spectrum for each of
the final states considered, along with the corresponding $p$-values. We can see that the $hh$ spectrum is very
similar to the $b\bar{b}$ spectrum, while the spectra from $(W^+W^-,
t\bar{t}, ZZ)$ are somewhat harder. This difference in spectrum shape leads to the slightly better fit for Higgs with respect to the other final states.

\subsection{The Fermi Analysis}

Very recently the Fermi collaboration has also weighed in on the
GCE~\cite{simonatalk}. Due to the preliminary nature of their result
not all of the information regarding the analysis is presently
available. However, it is known that an additional contribution with
an NFW morphology improves the fit to the Galactic center data
significantly. Fermi has shown the best fit energy spectra which fit
the excess, for four qualitatively different background models which
we dub Fermi spectra (a)-(d), and which are shown in
Figure~\ref{fig:at-rest}. The span of these spectra is significantly
larger than the uncertainties derived 
by~\cite{Calore:2014xka} and is thus a more conservative measure of the systematic uncertainties.

The fit uncertainties associated with the best-fit spectra are not yet
presented and thus we cannot do a proper fit to the data. Nonetheless,
we can use the Fermi spectra to locate regions in the
$m_\chi$--$\langle \sigma v \rangle_\mathrm{eff}$ plane which
reproduce the Fermi best fit spectra. Doing this will \emph{not} give
us a correct estimate of the expected fit uncertainty. \emph{It will,
however, give us an idea as to the range of possibilities that the
Fermi spectra allow, given the systematic uncertainties}.

Since we are not truly fitting to the Fermi data we have several
choices about how to present our result which we will now describe.
The analysis carried out by Fermi made two different assumptions about
the spectrum of the GCE.  In one case they simply let the
normalization of the excess float in each energy bin, in the other
they assumed the spectrum had the form of an exponentially cut-off
power law, i.e. $E^2 dN/dE = N_0 E^{-\gamma} e^{-E/E_c}$.  The final
states we consider are remarkably well fit by this simple form and so
we choose to use the GCE spectra Fermi derived under this assumption.  
Also, there are various possible ways to determine the best fit region
in the $m-\langle\sigma v \rangle_\mathrm{eff}$ plane \eg\ one could
allow any spectrum which fits in the envelope between the 4 presented
spectra, or one could fit each spectrum separately to get a feel for
the systematic uncertainty. Here, we take the latter approach.

Out of the 4 spectra Fermi (a,b,c,d) present, one (a) has a shape very different from that of heavy DM annihilating to electroweak final states.  Furthermore, fitting to (a) gives results similar, although with smaller allowed regions, to the CCW fit presented above.  Also, it is quite possible that spectrum (a) will be compatible with our spectrum shapes once the Fermi fit uncertainties are released.  Of the other spectra, (c) and (d) are very similar.  We therefore choose to present results for two of the spectra presented (b and d) in \cite{simonatalk}. Each of the spectra will give us a preferred region in the $m_\chi$--$\langle \sigma v \rangle_\mathrm{eff}$ plane. The systematic uncertainty is then quantified by ``sliding'' one preferred region to the other.

In order to fit each individual spectrum we must estimate its associated uncertainties. We take this uncertainty to be the statistical uncertainty on the total event rate in every energy bin (also given in~\cite{simonatalk}). Making a different choice (or, for that matter, using the correct uncertainty, once available) will only change the size of each allowed region, but will not change the off-set between the two allowed regions, which quantifies the systematic uncertainty and which is the main point of this exercise. With this choice we can now carry out a $\chi^2$ fit to the Fermi spectra, but we point out that because we are fitting to a best fit spectrum with a prescribed functional form, the values of the ``$\chi^2$'' will not carry rigorous statistical significance.

As for the the DM profile and the resulting uncertainties on $\mathcal{J}$, the Fermi analysis considers a region of $15^\circ\times 15^\circ$ around the Galactic center, smaller than that of CCW. However the Fermi analysis does not mask-out the Galactic center.  It was performed for NFW profiles with slopes of 1.0 and 1.2, finding good fits for both (with similar spectra). The larger uncertainty in $\gamma$ is to be expected because that analysis does not extend as far out from the Galactic center and thus has a smaller ``lever-arm''.
To be conservative, we will use the more constrained range of $\gamma\sim 1.2\pm 0.1$ from the CCW analysis in presenting our results. 
This choice leads to a line-of-sight integral of $\bar J_\mathrm{canonical}=1.58 \times10^{24}$ GeV$^2$/cm$^5$. Once we include the $\rho_\odot=0.4\pm0.2$ GeV/cm$^3$ uncertainty we get a range for the line-of-sight integral of $\mathcal{J}\in [0.14,4.0]$.
%Once a complete fit to the Fermi spectrum is done we expect this range to grow somewhat which would lead to a larger uncertainty band for the thermal relic cross section.

\begin{figure}[t]
  \begin{center}
    \includegraphics[width=0.8\textwidth]{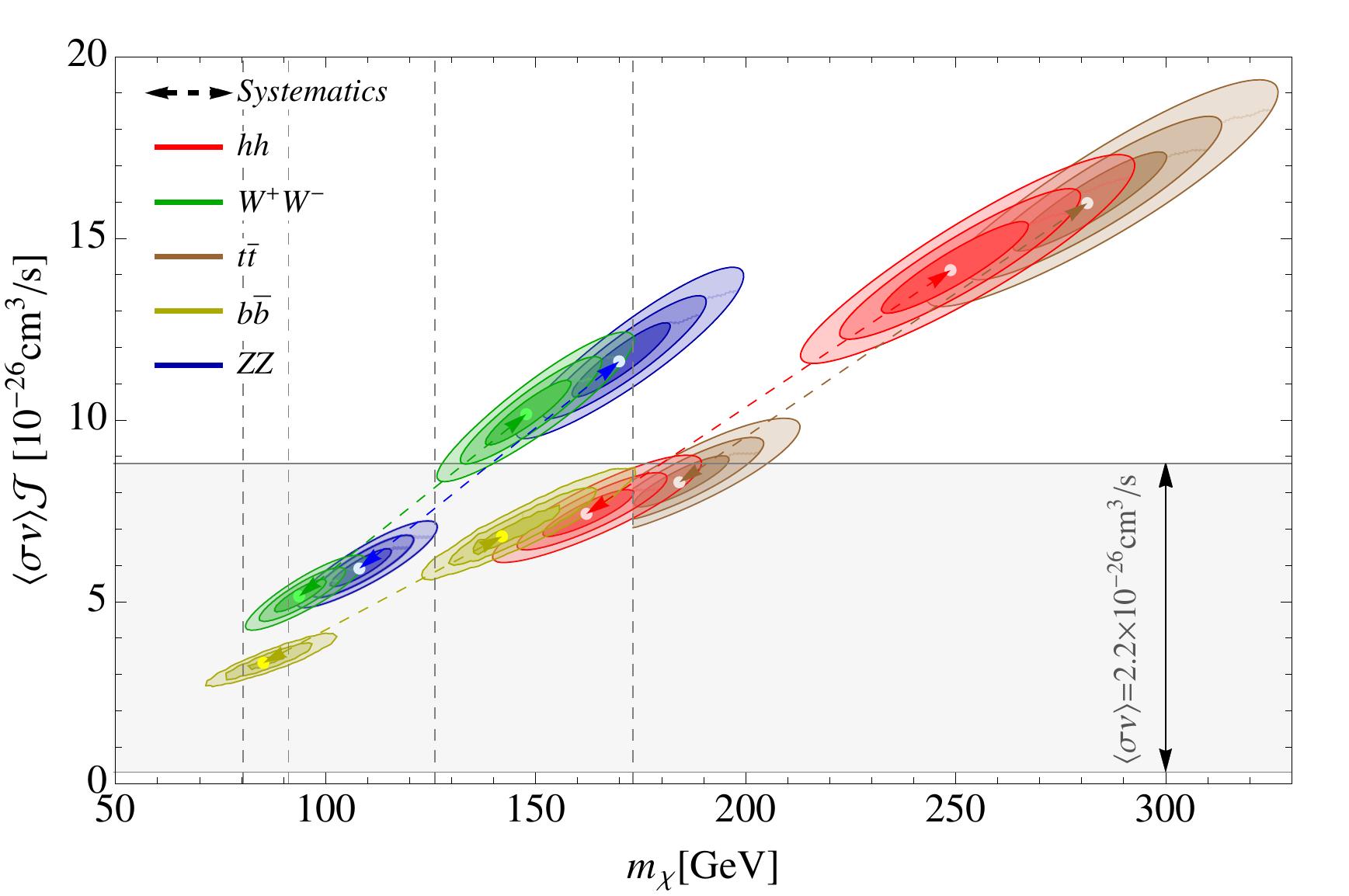}
    \\
    \includegraphics[width=0.485\textwidth]{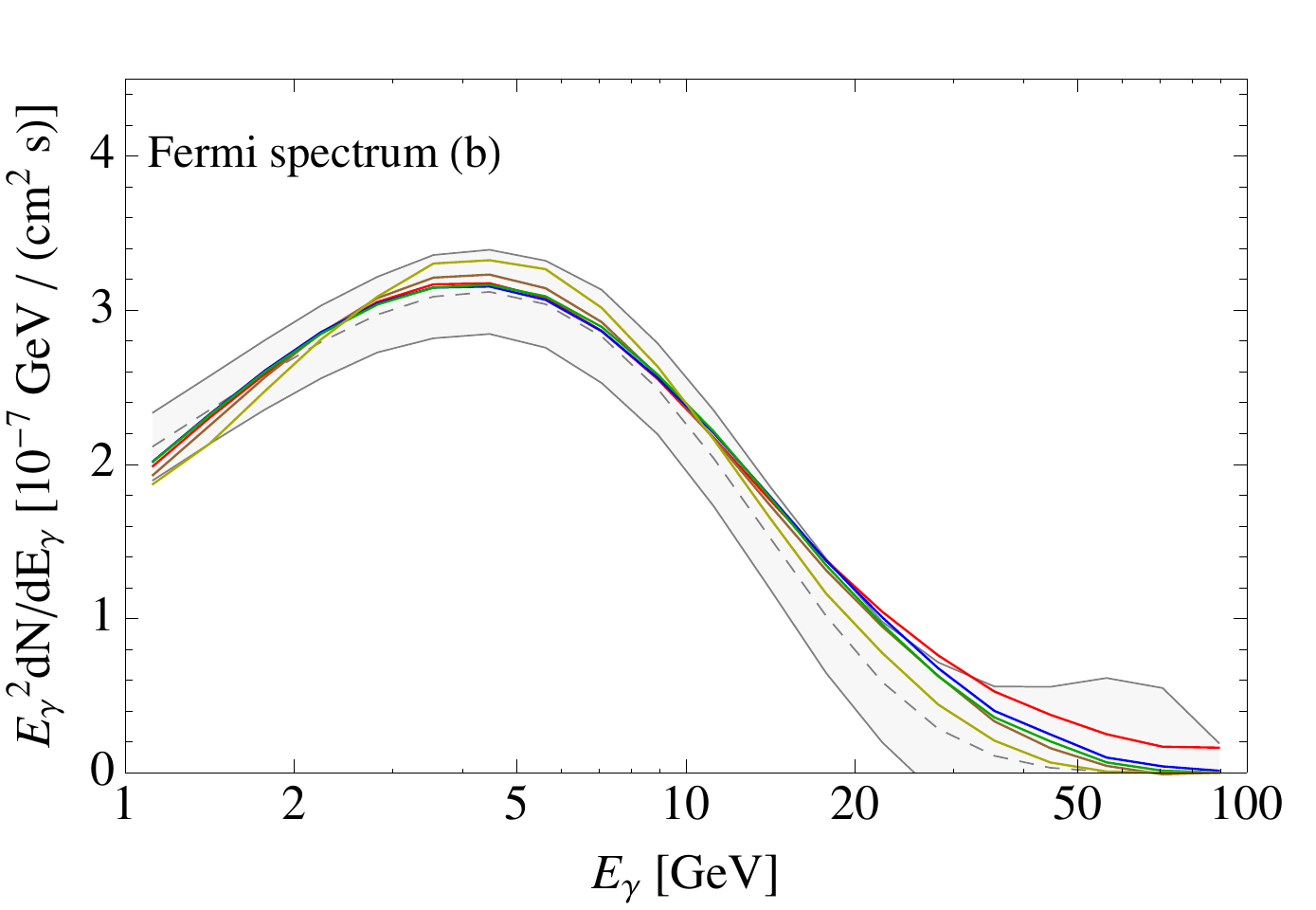}
    \includegraphics[width=0.485\textwidth]{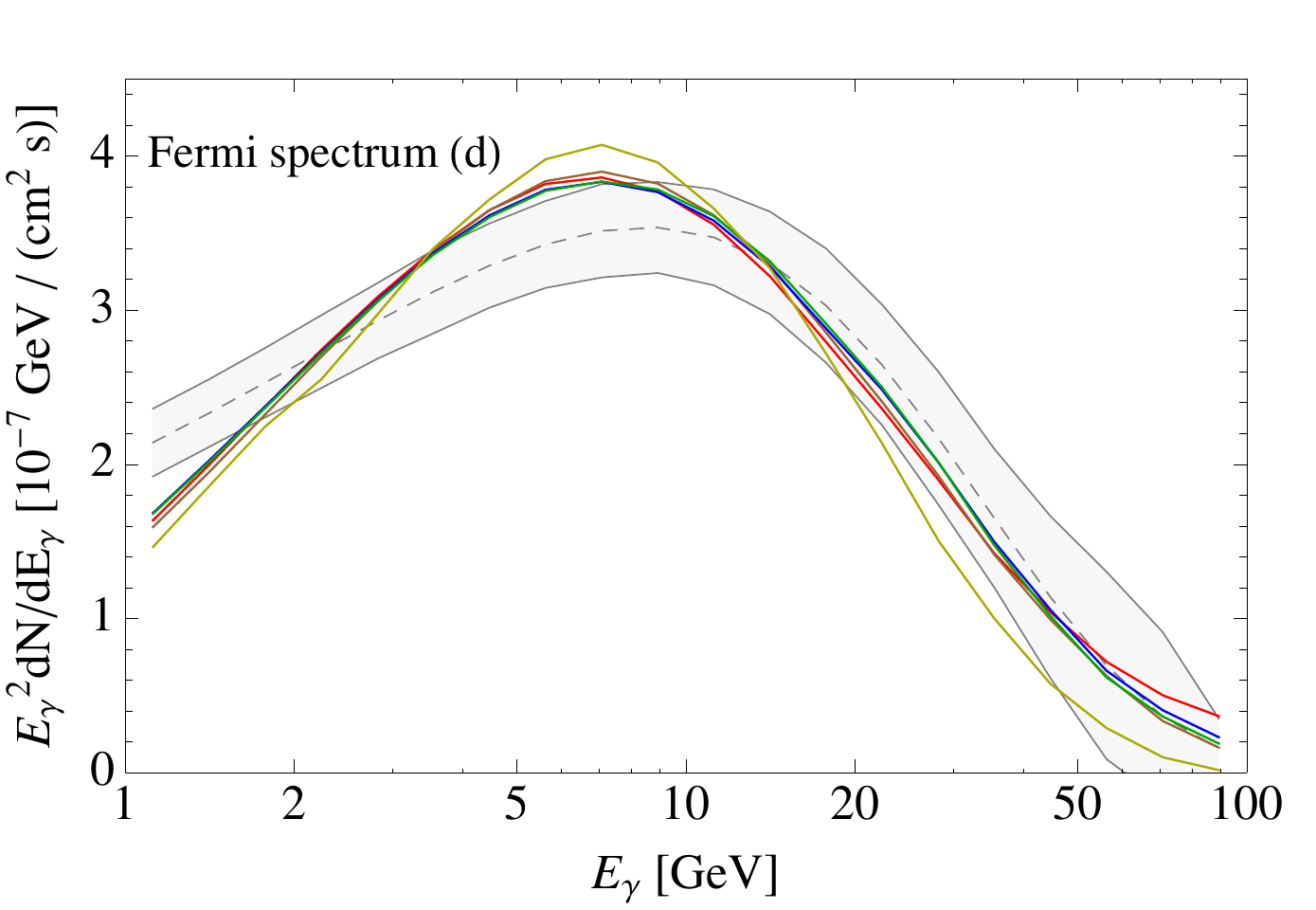}
  \end{center}
  \caption{\emph{Top:} Regions of parameter space which reproduce the Fermi best fit spectra.
  We do not fit to Fermi data, but rather to their reported best-fit spectra with statistical uncertainties only. 
  We show the ``$\Delta\chi^2$'' contours obtained for the hypotheses
  $\chi \chi \to X X$ for $X=\{h, W^\pm, Z, t, b\}$ fitting to Fermi's spectrum (a) (low mass) and spectrum (d) (high mass).
  Uncertainties from a full fit are likely to grow.
  Parameter space that is between the best fit regions, along the diagonal dashed lines, are also likely allowed by variations of the background model.
  \emph{Bottom:} 
  We show the spectra of photons obtained for the
  corresponding best fit values in the upper plot. Fermi spectrum (b) is on the left and spectrum (d) is on the right. The Fermi spectra are shown as a dashed line and the gray envelope shows the statistical uncertainty we used in the fits.
  }
  \label{fig:Fermi-money}
  \label{fig:money-Fermi}
\end{figure}

The results of this fitting procedure are shown in the top panel of Figure~\ref{fig:Fermi-money}, with the lower (higher) mass region in each case corresponding to the fit to spectrum b (d).  We have only studied Fermi spectra (b) and (d) but the region that lies between them is also viable and thus we expect that the region of parameter space that lies between the two fit regions to also provide a good fit, we denote this in the plot with dashed arrow-lines.  Notice that the low and high mass regions are also separated in cross section, and the cross section needed to fit the GCE scales as $\sim 1/m_\chi$.  This is as expected given the behavior seen in Figure~\ref{fig:progspec}, since, although the DM number density scales as $\sim1/m_\chi^2$, the number of photons per annihilation scales as $m_\chi$.

These results are qualitatively similar to the CCW fits, Figure~\ref{fig:chisquaredonshell}.  The result reinforces our conclusion from the CCW fit and takes them further: once uncertainties are taken into account, the allowed region extends to higher DM masses than previously realised, and allows for a greater range of final states. 
The growth in the range of masses brought about by the Fermi analysis is particularly striking for the $b\bar b$ final state which was in the neighborhood of 30-60 GeV in most previous studies.  One may consider the union of the regions allowed by the CCW and the Fermi studies as spanning the range of possibilities, and the large size simply reflects our present lack of understanding of the backgrounds and thus on the exact form of the excess. This expanded range of possibilities and new set of final states has a significant impact on the range of particle physics models that can explain the GCE. We will explore some examples in Section~\ref{sec:models}.

The bottom panels of Figure~\ref{fig:Fermi-money} show the best fit spectra for each of our channels compared to Fermi spectra (b) and (d) (on the left and right respectively) along with the statistical uncertainties we took for the fit. We see that the Fermi power-law-with-cutoff parametrization can be matched by many well motivated particle physics models.  For spectrum (b) the fits are remarkably good, for the best fit points in $(b\bar{b}\,,W^\pm W^\mp\,,ZZ\,,hh\,,t\bar{t})$ final state the $\chi^2$, for the 20 bins of the Fermi result, are $(2.6\,,1.8\,,2.6\,,4.6\,,2.0)$.  For spectrum (d) the corresponding $\chi^2$ are $(44\,,15\,,15\,,20\,,21)$.

\subsection{Considering Multi-Component Dark Matter  -- $f_\chi<1$}\label{sec:wtf}

So far we have been considering the case where all of the DM contributes to the GCE. It is also possible that the DM that is responsible for the GCE is not the only species of DM in the Galaxy, so called multi-component DM \ie\ $f_\chi<1$.  We now address the question of how little of the DM need be in this component to still give a sufficiently large photon signal. 

As the fraction of DM involved in producing photons drops the annihilation cross section must grow, such that $f_\chi^2 \langle \sigma v\rangle\mathcal{J}$ is approximately constant.  At the same time if DM is a thermal relic its abundance is set by the annihilation cross section and $f_\chi\sim 1/\langle \sigma v\rangle$.  As can be seen in Figures~\ref{fig:chisquaredonshell} and \ref{fig:Fermi-money} there is some overlap, due to the uncertainty in $\mathcal{J}$, between the $f_\chi=1$ thermal cross section and the cross section required by the signal.  As the cross section increases the amount of (thermal) DM drops and at some point the abundance will be too small to explain the GCE. 

\begin{figure}[tp]
  \begin{center}
    \includegraphics[width=0.48\textwidth]{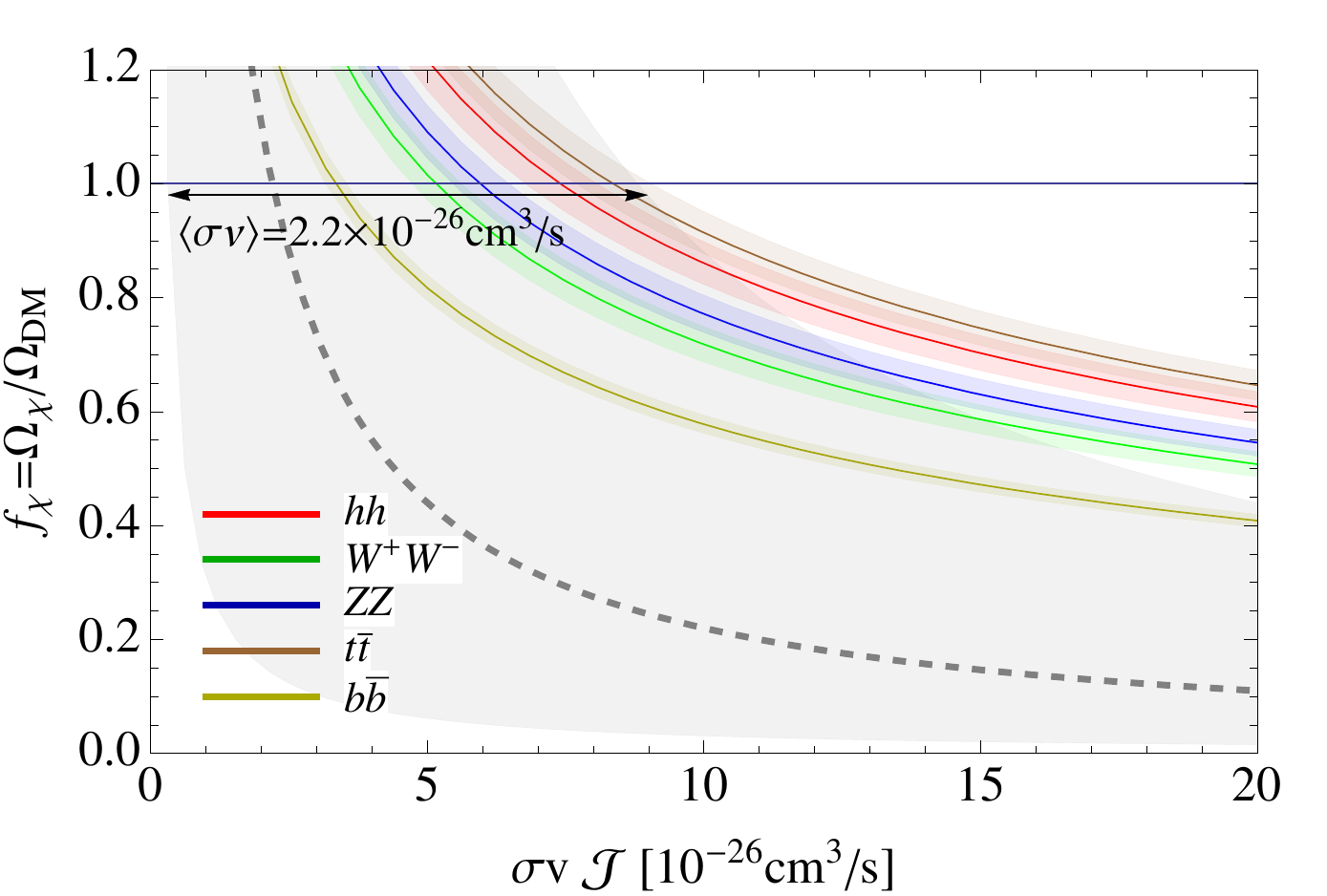}
\quad
    \includegraphics[width=0.48\textwidth]{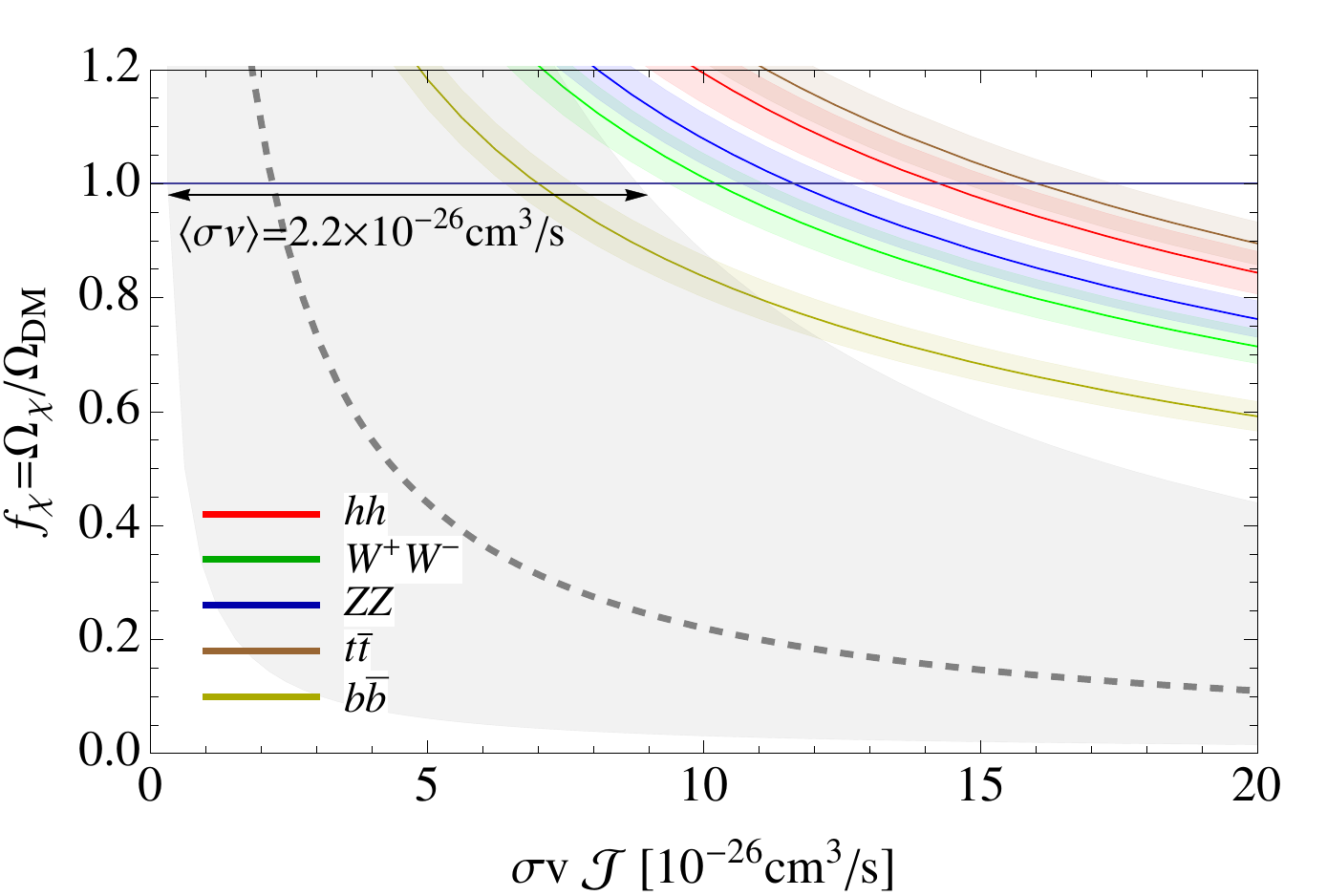}
  \end{center}
  \caption{The fraction, $f_\chi$, of the relevant component of dark
matter as a function of its annihilation cross section, including the
uncertainty in $\mathcal{J}$. Colored bands correspond to 1$\sigma$
preferred regions for the Galactic center excess. The gray band indicates the region where
the dark matter is populated thermally. On the left (right) the regions are shown for the fit to Fermi spectrum b  (d).}
  \label{fig:fractions}
\end{figure}

In Figure \ref{fig:fractions}, we show the deviation from $f_\chi=1$. The various colored bands are the regions that can explain the Fermi excess, where $f_\chi^2 \langle \sigma v \rangle \mathcal{J}$ lies within the ``$1\sigma$'' contours of the Fermi fits, the left-hand plot is for spectrum (b) and the right-hand plot is for spectrum (d). The dashed gray line shows how the thermal abundance changes as $\langle \sigma v \rangle$ varies, assuming $\mathcal{J}=1$ and the band around it encapsulates the uncertainty, \ie\ $\mathcal{J}\in [0.14,4.0]$.  Regions where the colored bands overlap the gray band show regions where the DM is not all of the DM, but can be a thermal relic, and can explain the GCE.  We see that for $b\bar{b}$ both spectrum (b) and (d) can have a thermal, but $f_\chi<1$, explanation.  For other final states only spectrum (b) is open to this interpretation.

An alternative possibility is that DM is not a thermal relic, its abundance is set by other means \eg\ late decay of another particle.  For this case the colored bands show that there is a large region of parameter space where the current signal can be obtained, for both spectra.   In Section \ref{sec:models} we
will present very simple models which realize this possibility.

This discussion has assumed that the annihilation cross section in the early universe, which determines the relic abundance, and the annihilation cross section now, which determines the photon flux, are the same.  If this is not the case the gray band will move.  For instance if there is co-annihilation in the early universe, the gray band will shift down, but if the annihilation is Sommerfeld enhanced the band will shift up.

%%%%%%%%%%%%%%%%%%%%%
%%%%%%%%%%%%%%%%%%%%%%
\begin{comment}
\begin{figure}[tp]
  \begin{center}
    \includegraphics[width=0.7\textwidth]{figs/fluctuations.pdf}
  \end{center}
  \caption{We show here the nature of fluctuations in
  \cite{Calore:2014xka}. The central values are denoted by the orange
  circles. We have included some representative fluctuations of the
  data as solid curves, each of which is within $1\sigma$ of the data.
  We see that the correlated error bars make a visual comparison of
  the fit hard.}
  \label{fig:fluctuations}
\end{figure}
\end{comment}
%%%%%%%%%%%%%%%%%%%%%%%%
%%%%%%%%%%%%%%%%%%%%%%%%

\section{Models}\label{sec:models}

We have seen that dark matter annihilation to $W$'s, $Z$'s, $h$'s and
$t$'s can provide a good fit to the Gooperon excess in a large window
of masses as summarized by Table~\ref{tab:massrange}. We will now
consider some models of dark matter which annihilate to these final
states. 
The importance of investigating these models in addition to the more
studied low-mass $b\bar b$ final state is the new possibilities this
opens up for model building. For example, assuming the GCE comes from
$b \bar b$ it was concluded that the excess cannot be accommodated
within the MSSM and that the addition of a singlet and ten new
parameters was needed~\cite{Cahill-Rowley:2014ora}. Although we will not carry out a
comprehensive study, we shall see by
example that there are simple supersymmetric models able to account for the GCE with annihilations to $W$'s or
tops. In addition, we suspect that the enlarged window of mass opened
by the Fermi analysis allows for simple SUSY models with
annihilations into $b\bar b$.

We will consider several examples of dark matter models which produce these
annihilation channels.  For each model we will derive constraints
from direct detection experiments, high energy solar neutrino
searches, and collider experiments. We will also estimate the
parameters for which the
thermal relic density matches observation. The thermal relic region
will coincide with the preferred Gooperon region in some cases, but we
note that non-thermal dark matter production can be invoked when it
does not.   

In addition, there can be important constraints from correlated indirect detection signals. In particular, the recent studies of Ref.~\cite{Bringmann:2014lpa,Cirelli:2014lwa,Hooper:2014ysa} have examined the constraints from cosmic ray antiproton measurements on possible DM interpretations for the GCE (for the $b\bar b$ final state). Unfortunately these limits strongly depend on the assumptions made in the modeling of the cosmic ray propagation and solar modulation~\cite{Cirelli:2014lwa}, and reasonable choices exist which allow compatibility with a GCE explanation. We note the heavy final states that are the focus of our work will lead to even harder antiproton spectra, and thus could potentially provide a handle in distinguishing them from other DM interpretations. A complementary indirect probe comes from gamma-rays from DM-dominated dwarf spheroidal galaxies. The limits on DM annihilation to hadronic final states are starting to probe the interesting regions favored 
by the GCE~\cite{dwarftalk}. We note that a careful analysis of the compatibility of the GCE and dwarf limits should take into account the uncertainties related to the dark matter halo, such as the ones discussed in Section~\ref{sec:halo uncertainties}. 
A dedicated study of these and other related indirect detection signals is warranted, which we leave to future work.

\subsection{The MSSM Neutralino}

We begin by studying the canonical example of a WIMP -- the neutralino of the MSSM.
Depending on its content (bino, wino, higgsino) and the presence of other light SUSY states in the spectrum, 
the neutralino can exhibit a variety of annihilation channels, including $WW$, $ZZ$, $hh$ and $t\bar t$ as well as $b\bar b$. Therefore, in addition to being a theoretically motivated WIMP candidate, 
the neutralino provides a versatile scenario in which to explore various final states from DM annihilation. 

We first consider the situation in which the electroweakinos are the only light supersymmetric states available and decouple all squarks, sleptons, and the second Higgs doublet.  In this limit, the relevant parameters are $\mu$, the higgsino mass parameter, the bino and wino mass parameters, $M_1$ and~$M_2$, as well as the ratio of the Higgs vevs, $\tan\beta$. We will begin by considering scenarios where all of the mass scales are $\mathcal{O}(100)$ GeV and the dark matter is an admixture of higgsinos and bino and wino. Later we will explore other options, including the case when the LSP is an almost pure state, as well as the effects of a light stop in the spectrum.

\subsubsection{Mixed Neutralino}\label{sec:neutralino}

Even within our simplified limit of the MSSM, the dark matter can exhibit a wide range of phenomenological possibilities. We begin here by considering the simple case of a mixed neutralino annihilating to electroweak gauge bosons. Here we will present a few sample slices of parameter space in which the mass and annihilation rate suggested by the GCE can be obtained while simultaneously satisfying the relevant phenomenological constraints. This is not intended to be exhaustive but rather meant to motivate more thorough studies. We will begin by assuming that all other particles are heavy. In this case $WW$ is often one of the dominant annihilation channels.

Our fits in Section~\ref{sec:fits} tell us that in order to describe the GCE a neutralino annihilating into $W$'s must have a mass between $m_W$ and 165 GeV, and an annihilation cross section in the Galaxy today that is close to the thermal value, $\langle \sigma v \rangle \approx 2.2 \times 10^{-26}\, {\rm cm}^3 \,{\rm s}^{-1}$. For such light LSPs,  the annihilation cross section of a  pure higgsino or wino is much too large, while that of a pure bino is too small (see section~\ref{sec:HiggsinoWino}). However, as is well-known, the appropriate cross section can be obtained if the neutralino is ``well-tempered''~\cite{ArkaniHamed:2006mb}, i.e., is a mixed state lying close to the boundary between a bino LSP and a wino or higgsino LSP. We will mostly focus on the case in which the LSP is primarily a  bino/wino mixture, $M_1 \sim M_2, \sim {\cal O}(100\, {\rm  GeV})$ and 
$|\mu| > {\cal O}(200 \, {\rm GeV})$, such that the dominant annihilation channel is indeed $WW$.  We reserve some comments for the bino/higgsino option for the end of this section.   

In the well-tempered regime, the splittings between the LSP and the next lightest states in the spectrum are typically small, and therefore co-annihilation processes are active in the early universe. This implies that the effective annihilation cross section governing thermal freezeout is always larger than the late-time annihilation cross section relevant for the GCE. Thus, we typically expect a slight mismatch between regions favored by the GCE and those suggested by a thermal relic. In our results below, we will keep track of the relic abundance calculation, paying attention to whether the thermal abundance overlaps with the region that fits the GCE. We note however that the abundance of dark matter could be set by a non-thermal mechanism, so models that do not thermally produce the observed DM abundance are by no means ruled out. 

\begin{figure}[t] %  figure placement: here, top, bottom, or page
   \centering
   \includegraphics[width=0.48\columnwidth]{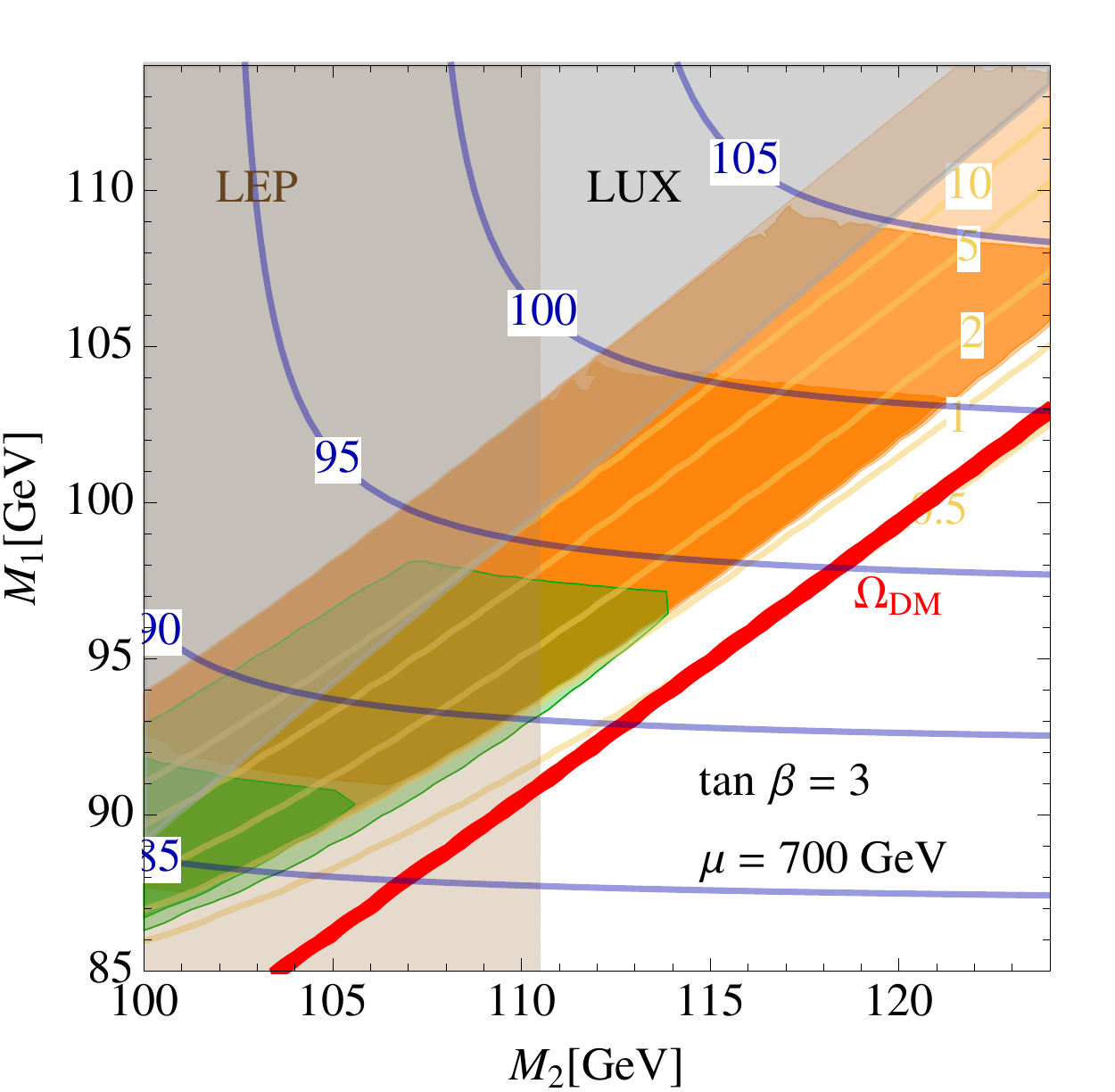} 
   \quad
      \includegraphics[width=0.48\columnwidth]{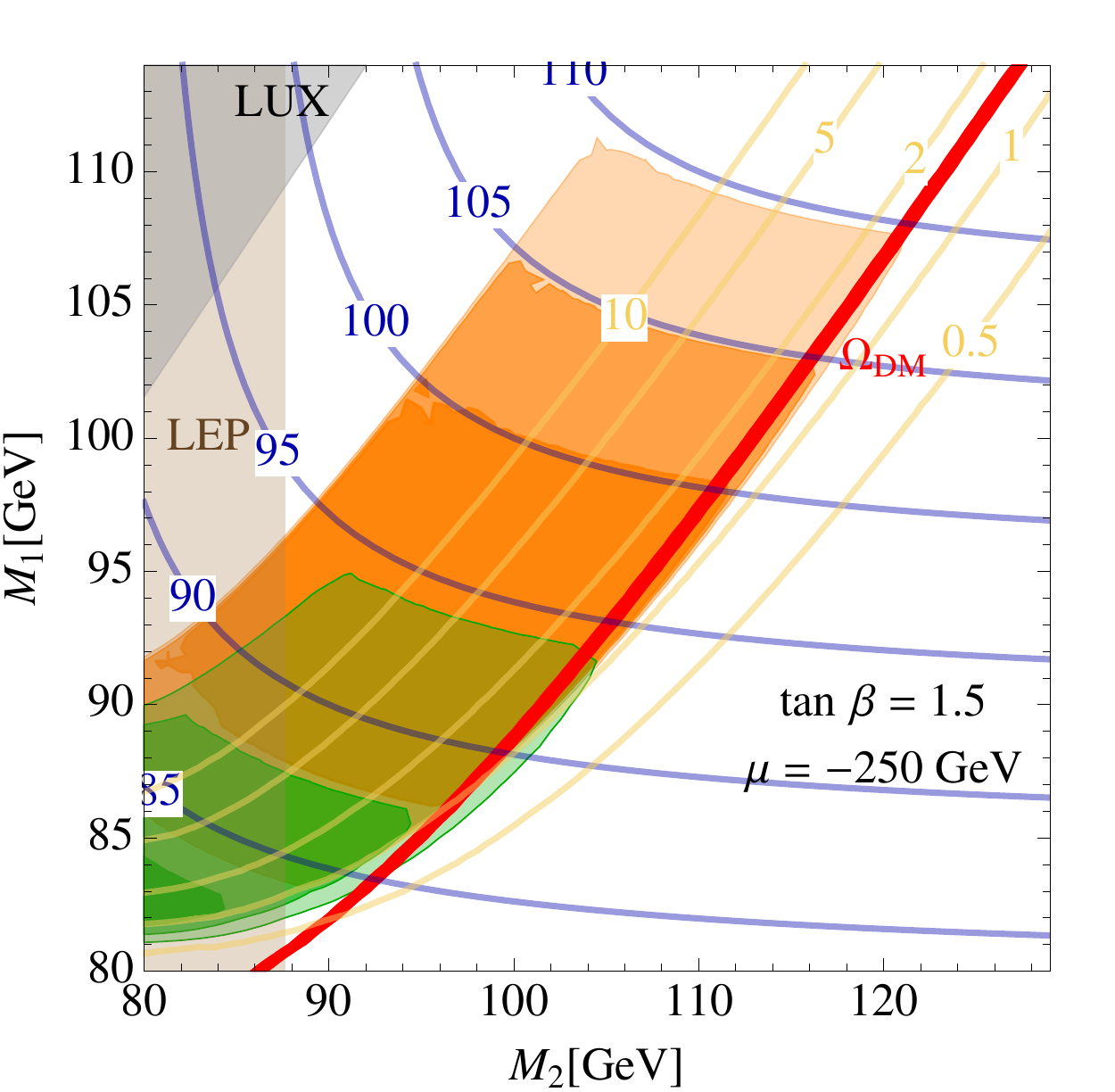} 
   %\qquad
%   \includegraphics[width=0.45\columnwidth]{figs/MSSM4.pdf} 
   \caption{
   %\draftnote{Pick only some of these, which ones?}
  {\it Mixed MSSM neutralino}. We display the 1,2,3$\sigma$ best-fit GCE regions for the $WW$ final state in the $M_2 - M_1$ plane for CCW (green) and Fermi spectrum (b) (orange). 
 We also overlay constraints from LEP chargino searches (brown) and LUX (gray).  In the (red) region denoted $\Omega_{\rm DM}$ the thermal relic abundance for the DM is within $3\sigma$ of the observed value.
For convenience, we also show the mass of the DM and the annihilation cross section to $WW$ as blue and yellow contours in units of GeV and  $10^{-26}$~cm$^2/s$ respectively.   In the left plot we have fixed $\mu = 700$ GeV, $\tan\beta = 3$, while in the right plot we have fixed 
 $\mu = -250$ GeV, $\tan\beta = 1.5$.
%\draftnote{$@$Brian: you mentioned another possible plot here, was this it?}
}
   \label{fig:MSSM}
\end{figure}

To compute the neutralino cross sections, relic abundance, and light electroweakino spectrum we have used the MSSM implementation of the {\tt MicrOMEGAs} package~\cite{Belanger:2013oya,Belanger:2004yn,Belanger:2001fz} and the {\tt SuSpect} package~\cite{Djouadi:2002ze}, with all additional SUSY states decoupled. These results were cross checked with a private code.
In Figure~\ref{fig:MSSM} we show two example regions that can account for the GCE. The best-fit regions of the GCE for the $WW$ final state as discussed in Section~\ref{sec:fits}, marginalized appropriately over ${\cal J}$, are shown for both the CCW analysis (green) and Fermi spectrum (b) (orange). We reiterate however, that this later region can ``slide'' up to higher masses as shown by the systematic uncertainties in Figure~\ref{fig:Fermi-money}.
In both plots the (red) region denoted $\Omega_{\rm DM}$ predicts a thermal relic abundance for the LSP within $3\sigma$ of the preferred value, \ie\ $\Omega_{\rm DM} h^2 = 0.1198\pm3\times 0.0026$~\cite{Ade:2013zuv}. We have also overlaid the limits on the spin-independent WIMP-nucleon cross section from LUX~\cite{Akerib:2013tjd} and the limits from LEP chargino searches ($m_{\chi^\pm}\gtap103.5$ GeV)~\cite{charginobigsplit}. In the left (right) plot, we have fixed $\mu = 700\, (-250)$ GeV and $\tan \beta  = 3(1.5)$.

In both examples we find that a significant range of parameters can account for the GCE while obeying the LEP and LUX constraints. Moreover, in the right plot of Figure~\ref{fig:MSSM} we observe a small region where the thermal relic abundance agrees with observation and furthermore overlaps with the region favored by the fit to the GCE. This is in contrast to the left plot where the two regions are separated, and the GCE explanation requires DM to be produced non-thermally. This difference can be understood as a consequence of the mass splittings between the LSP and the NLSP in the regions favored by the GCE. Smaller values of $|\mu|$ and $\tan\beta$, such as those in the right plot, tend to increase the mass splitting between the LSP and NLSP. This in turn reduces the effect of co-annihilation, allowing the GCE and thermal relic regions to overlap. In fact, for negative values of $\mu$ the increase in the mass splitting is more pronounced while simultaneously the higgsino fraction of the LSP is further suppressed, allowing compatibility with the LUX bounds as is seen in Figure~\ref{fig:MSSM}.

The two examples in Figure~\ref{fig:MSSM}, while sampling only a small portion of the parameter space, provide existence  proofs that a light neutralino making up all of the DM, either thermally or non-thermally, can account for the GCE. They are also illustrative of the general issues at stake in a potential DM model interpretation of the GCE. In addition to the constraints examined above, the LHC experiments are starting to extend limits on electroweakinos beyond the LEP limits in some cases, and we refer the reader to the recent study of Ref.~\cite{Martin:2014qra}. While the LHC limits cannot exclude a neutralino explanation of the GCE at this stage, it would be very interesting to examine the prospects for Run II; for recent work along these lines, see Refs.~\cite{Gori:2013ala,Jung:2013zya,Schwaller:2013baa,Han:2014kaa,Low:2014cba,Yu:2014mda,Bramante:2014dza,Cirelli:2014dsa}.

We have also explored, though not exhaustively, the case in which the LSP is primarily a bino/higgsino mixture. While one can find regions satisfying the GCE, there are generally competitive constraints from LUX due to the sizable LSP higgsino content. One exception is the direct detection ``blind spot'' where the neutralino-Higgs coupling vanishes
~\cite{Cheung:2012qy,Huang:2014xua}. However, in this region, there are further constraints from IceCube's search for solar neutrinos~\cite{Aartsen:2012kia}, which in turn limits the spin-dependent scattering cross section mediated by the $Z$ boson, again challenging a GCE explanation. Keeping in mind that the systematic uncertainties in the GCE are likely even larger than we have estimated here, it would be worthwhile to explore the bino/higgsino option in a systematic fashion.

\subsubsection{Pure Winos and Higgsinos with $f_\chi<1$}\label{sec:HiggsinoWino}

We now consider the edges of the MSSM neutralino parameter space where the neutralino is an almost pure wino or almost pure higgsino. We will find that in these cases the neutralino can produce a good fit to the GCE if it is \emph{not} all of the dark matter and if it is abundance is set non-thermally. 

Consider first the case $M_2 \ll |\mu|, M_1$ such that the spectrum contains a light wino with nearly degenerate neutral and charged states ($\widetilde W^0$, $\widetilde W^\pm$). A pure wino LSP is predicted, for example, in models of anomaly mediation~\cite{Randall:1998uk, Giudice:1998xp}.
In the Galaxy today, the neutral $\widetilde W^0$ annihilates to $W$ bosons via $\widetilde W^\pm$ exchange. %To provide the best description of the gamma-ray excess, the wino should be close in mass to the $W$-boson, $m_W \lesssim M_2 \lesssim 100$ GeV 
%(see the fit in Figure~\ref{fig:chisquaredonshell}). 
The annihilation cross section is given by 
\begin{eqnarray}
\langle \sigma v\rangle^{\widetilde W}_{WW} & = & \frac{g^4}{8 \pi M_2^2}\frac{(1-m_W^2/M_2^2)^{3/2}}{(1-m_W^2/2 M_2^2)^2} . 
\label{eq:ann-pure-wino}
\end{eqnarray}
The tree-level formula above is quite accurate for such low mass winos, with corrections at the few percent level coming from higher-order electroweak and Sommerfeld effects~\cite{Hryczuk:2011vi,Bauer:2014ula, Ovanesyan:2014fwa, Baumgart:2014vma}. 
The annihilation for such light winos is very efficient, with a cross section (\ref{eq:ann-pure-wino}) of order $4 \times 10^{-24}\,{\rm cm}^3\,{\rm s}^{-1}$. 
Thus, if the wino is populated thermally it will make up only a small fraction of the dark matter. 

In fact, once co-annihilation effects are considered the effective annihilation rate in the early Universe is even larger. Including co-annihilation and using {\tt MicrOMEGAs} we find that for a 95 GeV mass the relic density of winos is $\Omega_{\tilde W}h^2\sim 2.5\times 10^{-4}$ and the predicted gamma-ray signal will be much too faint to explain the GC excess. On the other hand, a non-thermal wino that saturates the dark matter density will yield a much larger gamma-ray signal than that observed by Fermi and is thus excluded~\cite{Cohen:2013ama,Fan:2013faa}. However, there is a third possibility. 

\begin{figure}[tp]
  \begin{center}
    \includegraphics[width=0.6\textwidth]{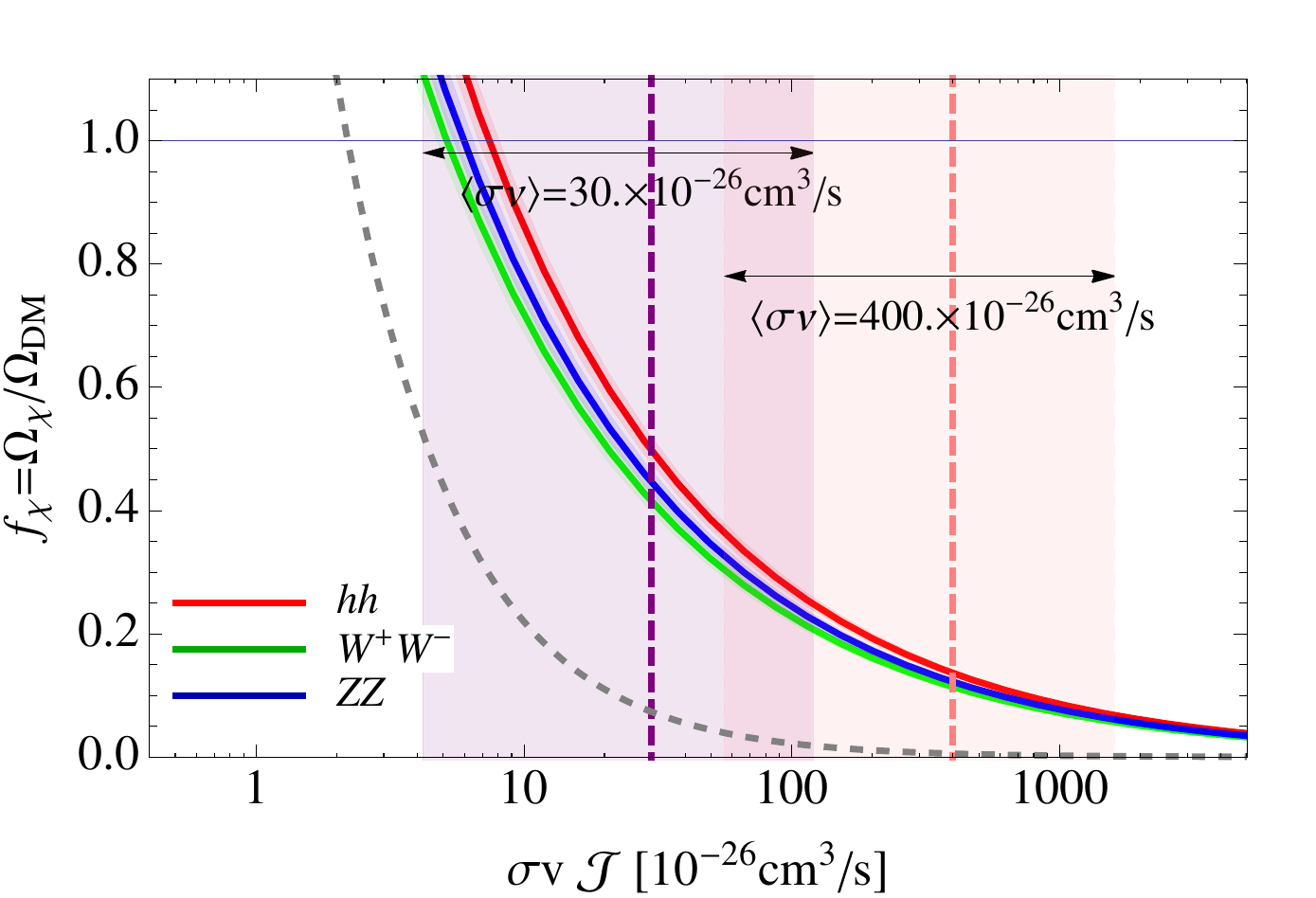}
  \end{center}
  \caption{
An extension of Figure~\ref{fig:fractions} to higher values of the effective cross section which include the expected cross sections of a higgsino (purple) and wino (pink). We show regions where a pure higgsino or wino can explain the GCE
provided they are populated non-thermally, and make up
$\mathcal{O}$(10\%) of the total dark matter.}
  \label{fig:fractionHiggsinoWino}
\end{figure}

Since the gamma-ray flux, equations (\ref{eq:flux}) and (\ref{eq:flux-J}), scales as 
$ \Phi \propto f^2_{\widetilde W^0} \langle \sigma v\rangle_{\widetilde W^0} $, it follows that a wino can still explain the gamma-ray signal if it is populated non-thermally with just a sub-dominant DM fraction $f_{\widetilde W^0} \approx \sqrt{\langle \sigma v \rangle^{\rm fit}_{WW}/\langle \sigma v\rangle_{\widetilde W^0} } \approx 0.1$, where $\langle \sigma v \rangle^{\rm fit}_{WW} \sim 4 \times 10^{-26}\,{\rm cm}^3\,{\rm s}^{-1}$ is the value inferred from, for example, the CCW fit (see Figure~\ref{fig:chisquaredonshell}) for the $WW$ channel assuming the DM saturates the observed relic density. The non-thermal origin of the wino could, for example come from the decay of moduli (as reviewed, for example, in~\cite{Fan:2013faa}). We show the wino dark matter region in the $\langle \sigma v\rangle\mathcal{J}$--$f_\chi$ plane in Figure~\ref{fig:fractionHiggsinoWino}.

The most important constraints on a light wino come from from collider experiments. These bounds depend sensitively on the mass splitting between the charged and neutral states, $\Delta M = M_{\widetilde W^\pm}-M_{\widetilde W^0}$. If tree-level mixing is negligible, $\mu \gg M_2$, there is still an irreducible radiatively generated  mass splitting $\Delta M \approx 150$ MeV (for $M_2 \approx 100$ GeV), which has been computed to two-loop order in Ref.~\cite{Ibe:2012sx}. For such a small splitting, the lifetime of the charged state $\widetilde W^\pm$ is of order $c \tau_{\widetilde W^\pm} \approx 10\, {\rm cm}$, such that production of $\widetilde W^\pm$ at colliders  will lead to the signature of a disappearing charged track. An ATLAS search for this signature excludes pure winos with radiatively generated mass splittings for $M_2 < 270$ GeV~\cite{Aad:2013yna}. However, if one arranges a slightly larger mass splitting, $\Delta M \gtrsim 220$ MeV by, for instance, a small mixing with the higgsino, then the ATLAS bound does not apply since the $\widetilde \chi^\pm$ lifetime is adequately shortened. Therefore, to evade the ATLAS disappearing track search, an additional $O(100 \,{\rm MeV})$ contribution to the chargino-neutralino mass splitting can be obtained through tree-level wino-higgsino mixing. In the limit $M_1 \gg |\mu| \gg M_2$, the tree-level contribution to the splitting is given by 
\begin{eqnarray}
m_{\chi^\pm} - m_{\chi^0} & \simeq & \frac{M_2 m_W^4 \cos^2{2\beta}} {2 (M_2^2-\mu^2)^2  }   \\
& \approx &  80\,{\rm MeV} \times \left( \frac{M_2}{100\,{\rm  GeV}} \right) \left( \frac{400\,{\rm  GeV}}{\mu}\right)^4, \nonumber 
\end{eqnarray}
where the second line is valid in the large-$\tan\beta$ limit. Thus, provided $|\mu| \lesssim 400$ GeV, this tree-level splitting, in tandem with the $\sim 150$ MeV radiative splitting from gauge boson loops, is sufficient to evade the ATLAS search. 

We also note that in this regime  of small  $O(100\,{\rm MeV} - 1\,{\rm GeV})$ mass splittings, the LEP limits are slightly weaker than  the naive kinematic limit, and neutralinos as light as 95 GeV are allowed~\cite{LEP2-chargino-DM}.  
Other searches, such as mono-jets and mono-photons, place no additional exclusions on the parameter space.

What about other constraints on this scenario? A pure wino has a spin-independent WIMP-nucleon scattering cross section $\sigma_p \sim 10^{-47}\,{\rm  cm}^2$~\cite{Hisano:2011cs,Hill:2011be}, which is currently unconstrained even if the wino comprises all of the dark matter. However, with additional tree-level mixing with the higgsino as discussed above, the neutralino obtains a sizable coupling to the Higgs boson, which leads to a cross section $\sigma_p \sim 10^{-45}\!-\!10^{-44} \,{\rm cm}^2$, in the range currently being probed by LUX. However, since we require a wino  comprising only a fraction $f_{\widetilde W^0} \approx 0.1$  of the dark matter, LUX does not currently constrain this scenario. 
Similarly, other indirect astrophysical probes that depend on the dark matter density, such as solar neutrinos, gamma-rays, or antiprotons from DM annihilation, do not constrain our scenario since the neutralino comprises a subdominant fraction of the total dark matter.

The case $|\mu| \ll M_1, M_2$ -  a ``pure'' higgsino\footnote{We assume a small mixing with the bino or wino such that the $Z$-boson mediated scattering with nuclei is inelastic~\cite{TuckerSmith:2001hy}.} - is very similar to the case of the wino just described. The lightest neutralino annihilates to $W$ and $Z$ bosons with cross sections given by
\begin{eqnarray}
\langle \sigma v \rangle^{\widetilde H}_{WW} & = & \frac{g^4}{128 \pi \mu^2}\frac{(1-m_W^2/\mu^2)^{3/2}}{(1-m_W^2/2 \mu^2)^2}, \nonumber  \\
\langle \sigma v \rangle^{\widetilde H}_{ZZ}    & = & \frac{g^4}{256 \pi c_W^4 \mu^2}\frac{(1-m_Z^2/\mu^2)^{3/2}}{(1-m_Z^2/2 \mu^2)^2} . 
\label{eq:ann-pure-higgsino}
\end{eqnarray}
For a 100 GeV higgsino, the total annihilation cross section is $\langle \sigma v \rangle^{\widetilde H} \approx 3 \times 10^{-25}\, {\rm cm}^3 \, {\rm s}^{-1}$. This is a factor of 10 to 25 to large to account for the GCE excess if the higgsino was all of the dark matter. However, due to coannihilation effects, the thermal abundance of the higgsino is suppressed, of order $\Omega h^2\sim3\times10^{-3}$. A thermal higgsino would thus contribute negligibly to the GCE.  As was the case for the wino, a non-thermal origin of higgsinos which comprise $\sim40\%$ of the total dark matter could account for the~GCE as shown in Figure~\ref{fig:fractionHiggsinoWino}.

In contrast to the case of a pure wino, the only collider limits on a pure higgsino come from LEP2. The ATLAS disappearing track search places no constraints on higgsinos. This is because 1) the production cross sections of higgsinos is smaller that of  winos by roughly a factor of two, and 2) the higgsino lifetime is shorter than the wino lifetime by roughly a factor of 6. The latter fact is due to the larger radiative splitting, which for a 100 GeV higgsino is about  $\Delta M  \approx 260$ MeV~\cite{Thomas:1998wy,Cirelli:2005uq}. 
The LEP2 combined limits allow for higgsinos as light as 92 GeV for mass splittings of order 200 MeV. 
No further constraints from direct or indirect dark matter searches exist for this scenario.

\subsubsection{Bino-Stop}
\begin{figure}[t] %  figure placement: here, top, bottom, or page
   \centering
   \includegraphics[width=0.6\columnwidth]{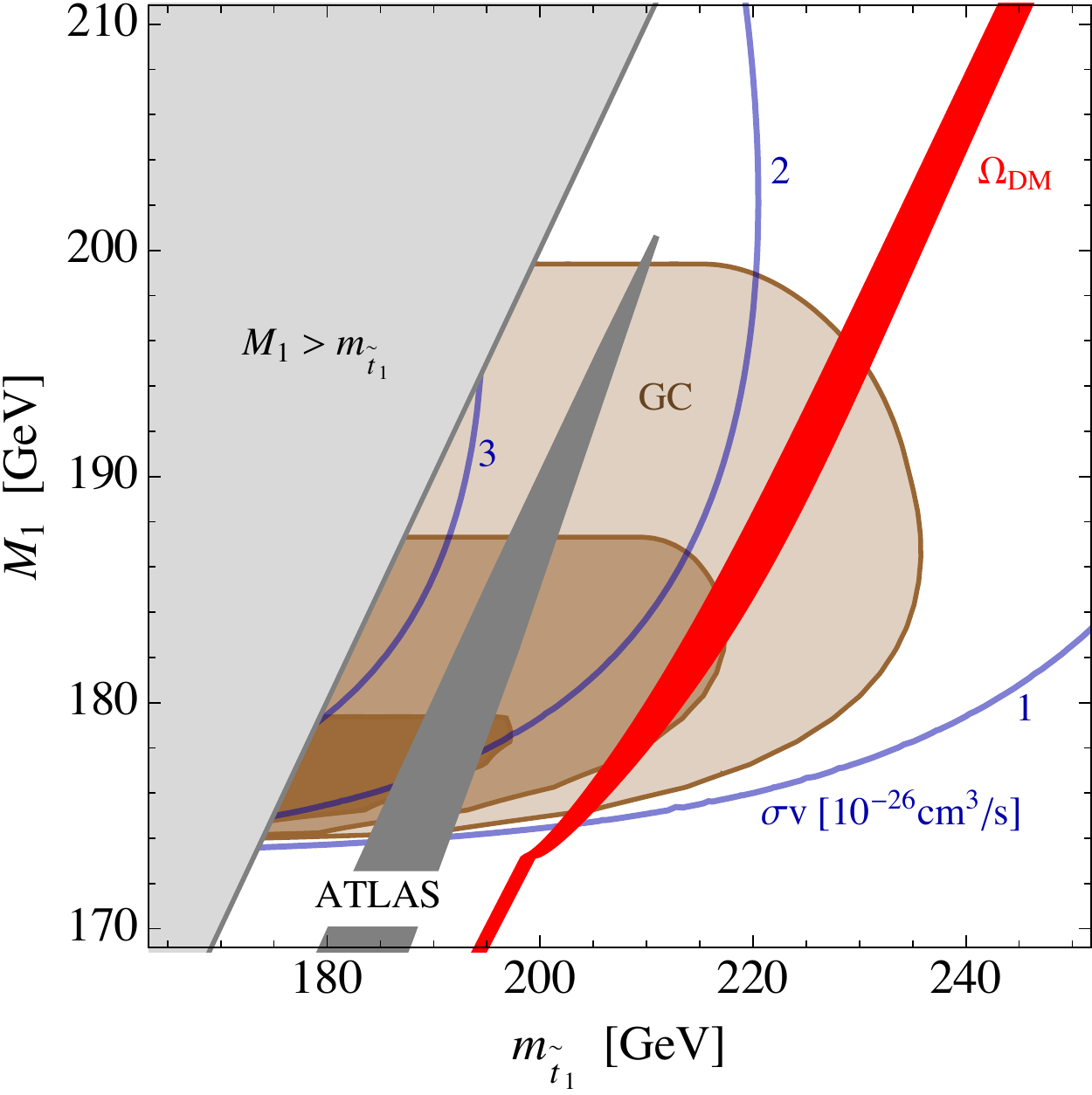} 
   \caption{{\it Bino-stop parameter space:} 
   We display the 1,2,3$\sigma$ $t\bar t$ best-fit regions for the CCW analysis of the GCE (brown). In the favored region, the annihilation cross section to top quarks lies in the range of $\langle \sigma v \rangle \approx (1-3) \times 10^{-26}\,{\rm cm}^3\, {\rm s}^{-1}$ (blue contours). We also display the region in which the bino is a thermal relic saturating the observed DM abundance (red), which overlaps with the GCE favored region. 
  Assuming a branching ratio ${{\rm{Br}}(\tilde t_1 \rightarrow c \chi^0 )} =  {{\rm{Br}}(\tilde t_1 \rightarrow b f f' \chi^0 )} = 50\%$, the ATLAS stop searches exclude the gray region~\cite{Grober:2014aha}.  
 }
 \label{fig:stop-bino}
\end{figure}

If there is a light top squark (stop) in the spectrum the neutralino can efficiently  annihilate to top quarks via $t$-channel stop exchange. Consider the simplifying limit in which the LSP is purely bino and the lightest stop is purely right-handed, $\tilde t_1  = \tilde t_R$, while the heavy left-handed stop is decoupled. The bino-stop-top coupling in this case is
\begin{equation}
{\cal L} = \frac{ 2 \, \sqrt{2} \, g'} {3} \tilde t_1 \, \tilde B \, \bar t +{\rm h.c.}, 
\end{equation}
and the annihilation cross section for $\tilde B \tilde B \rightarrow t \bar t$ is given by
\begin{equation}
\langle \sigma v \rangle = \frac{2 \, N_c \, g'^4 \, m_t^2}{81 \pi (M_1^2 + m^2_{\tilde t_1} - m_t^2)^2}\left(1 - \frac{m_t^2}{M_1^2}\right)^{1/2},
\label{eq:bino-stop}
\end{equation}
where $N_c = 3$. 
When both the bino and the stop are light, with masses of order $200 \, \rm GeV$, the annihilation cross section~(\ref{eq:bino-stop}) is in the range $\langle \sigma v \rangle \approx (1-3) \times 10^{-26}\,{\rm cm}^3\, {\rm s}^{-1}$. Given the uncertainties in the spectrum of the gamma-ray excess as well as ${\cal J}$, this scenario can yield an annihilation rate large enough to account for the GCE. In Figure~\ref{fig:stop-bino} we display the $t\bar t$ best-fit regions of the CCW analysis (Figure~\ref{fig:chisquaredonshell}) in the $m_{\tilde t_1} - M_1$ plane.

The cross section~(\ref{eq:bino-stop}) is required to be close to the canonical thermal relic value to explain the GCE. Since the bino-stop mass splitting is small, additional stop co-annihilation processes are important in the early universe~\cite{Boehm:1999bj,Ellis:2001nx,Balazs:2004bu,Balazs:2004ae}. 
We have estimated the bino relic abundance including the annihilation to $t\bar t$ via Eq. (\ref{eq:bino-stop}) as well as the effects of stop co-annihilation (including the Sommerfeld enhancement) using the results of Ref.~\cite{deSimone:2014pda}, cross checked with {\tt MicrOMEGAs}. In Figure~\ref{fig:stop-bino} we show the parameters for which a thermal relic bino makes up all of the dark matter. This region overlaps with a portion of the region explaining the GCE. Outside this region the bino must be non-thermally produced but still can explain the GCE. 

The most important constraints on this scenario come from direct searches for stops at the LHC. In this region of neutralino-stop mass parameter space, there are typically two stop decay channels that are considered: 1) the two-body flavor violating decay to charm and neutralino, $\tilde t_1 \rightarrow c \chi^0$, and 2) the four body off-shell decay $\tilde t_1 \rightarrow b f f' \chi^0$. ATLAS and CMS have performed searches optimized to these decay modes~\cite{Aad:2014kra,Aad:2014nra,CMS-PAS-SUS-13-009}, which, under the assumption that the branching ratio to either final state is 100$\%$, challenge a possible GCE explanation.

These limits are, however, very sensitive to the branching ratios of the two final states, and this dependence has recently been examined in detail in Ref.~\cite{Grober:2014aha}. For example, with a $\sim 50\%$ branching ratio to both the two- and four-body final states, there is still a significant region of stop--neutralino mass parameter space that is not covered by the ATLAS and CMS searches. From a low energy perspective, these branching ratios are essentially free parameters since they depend crucially on the effective stop--scharm mixing angle. This mixing angle can be large enough for the two branching ratios to be competitive without running afoul of flavor constraints; see e.g.~\cite{Agrawal:2013kha,Blanke:2013uia}. In addition, there have been several studies recasting existing analyses to constrain stops in this mass region~\cite{Agrawal:2013kha,Blanke:2013uia,Krizka:2012ah,deSimone:2014pda}, although again the limits depend crucially on the branching ratios and stop-neutralino mass splitting. Although beyond our current scope, it would be very interesting to perform a careful study of the existing limits and prospects for LHC Run-II, while taking into account the freedom in the stop branching ratio. Beyond these two decay modes, one can also contemplate a possible decay mode to $\tilde t_1 \rightarrow u \chi^0$, which could be probed by monojet searches. 

Beyond the constraints from direct searches at the LHC, there are indirect constraints on light stops from precision electroweak data, flavor physics, and Higgs signal strength measurements, but these depend on the details of the spectrum and mixings of other sparticles, and no firm bounds can be placed. Furthermore, direct detection experiments place no limits on this scenario due to the bino nature of the dark matter and its negligible coupling to quarks other than the top. 

\subsection{Higgs Portal Dark Matter}\label{sec:pospelon}

\begin{figure}[t] %  figure placement: here, top, bottom, or page
   \centering
   \includegraphics[width=0.7\columnwidth]{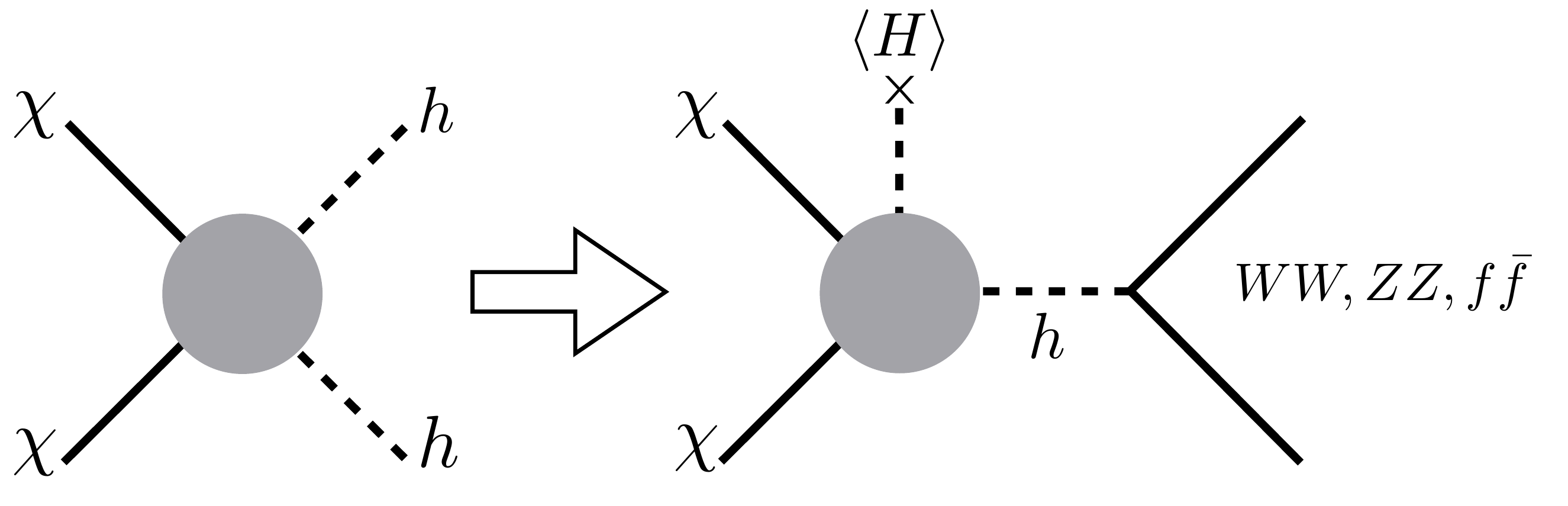} 
   \caption{Dark matter annihilation to Higgs bosons generally implies annihilation to other SM final states, and scattering at direct detection experiments.}
   \label{fig:hh}
\end{figure}

A well motivated class of WIMP dark matter models is the so-called Higgs portal dark matter, where the dark matter couples to the standard model via the Higgs boson. The minimal example of such a model consists of a new scalar singlet $\chi$ which
couples to the Higgs with a renormalizable interaction~\cite{Silveira:1985rk,McDonald:1993ex,Burgess:2000yq}
 
\begin{equation}
-{\cal L}\supset   \frac{1}{2}\lambda \, \chi \, \chi \, H^\dag H + \frac{1}{2}m^2  \chi \, \chi
%& \supset &  \frac{1}{2}\lambda v  \, \chi \, \chi \, h + \frac{1}{4}\lambda  \,  \chi \, \chi \, (h^2 + \varphi_0^2 + 2 |\varphi_+|^2).
\label{eq:higgs-portal}
\end{equation}
Naively this may be a good candidate for dark matter annihilating to Higgses, however, generically, and as illustrated in figure~\ref{fig:hh} annihilation into Higgses also implies annihilation into the other final states we consider, $W$'s, $Z$'s and tops (if the dark matter is above threshold).
For example, for a dark matter mass of $m_\chi=$170 GeV and a portal coupling $\lambda \lesssim 0.1$, the annihilation fraction into ($WW$ : $hh$ : $ZZ$) is roughly (51\% : 26\% : 23\%).  It is interesting to note that if we interpret the region preferred by Fermi, Figure~\ref{fig:money-Fermi}, quite loosely, there is significant overlap between the dark matter masses preferred by these final states. As we will now show, dark matter which annihilates to an order one admixture of these states could also fit the observed spectrum.

\begin{figure}[t] %  figure placement: here, top, bottom, or page
   \centering
   \includegraphics[width=0.485\columnwidth]{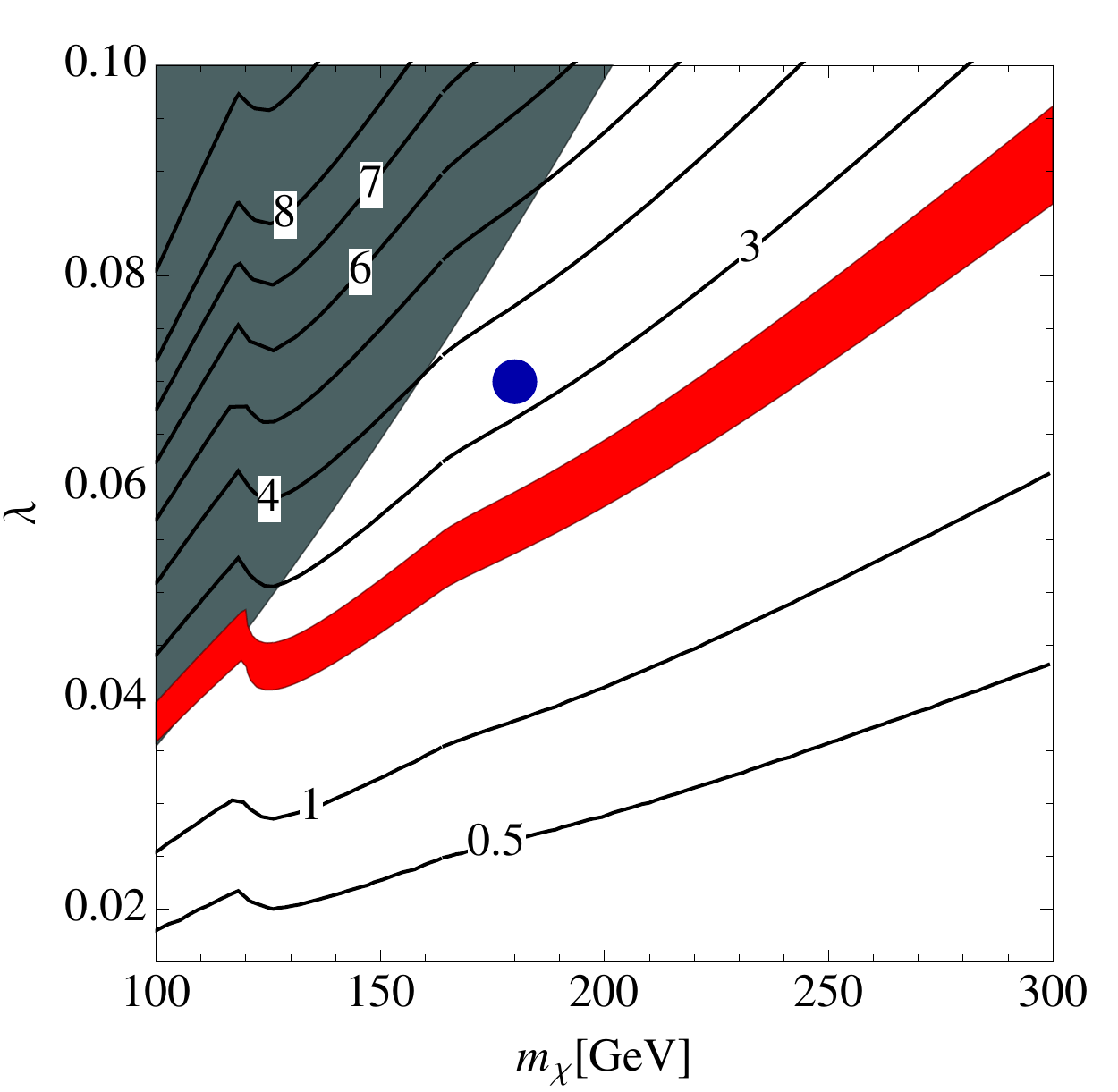} 
   \includegraphics[width=0.485\columnwidth]{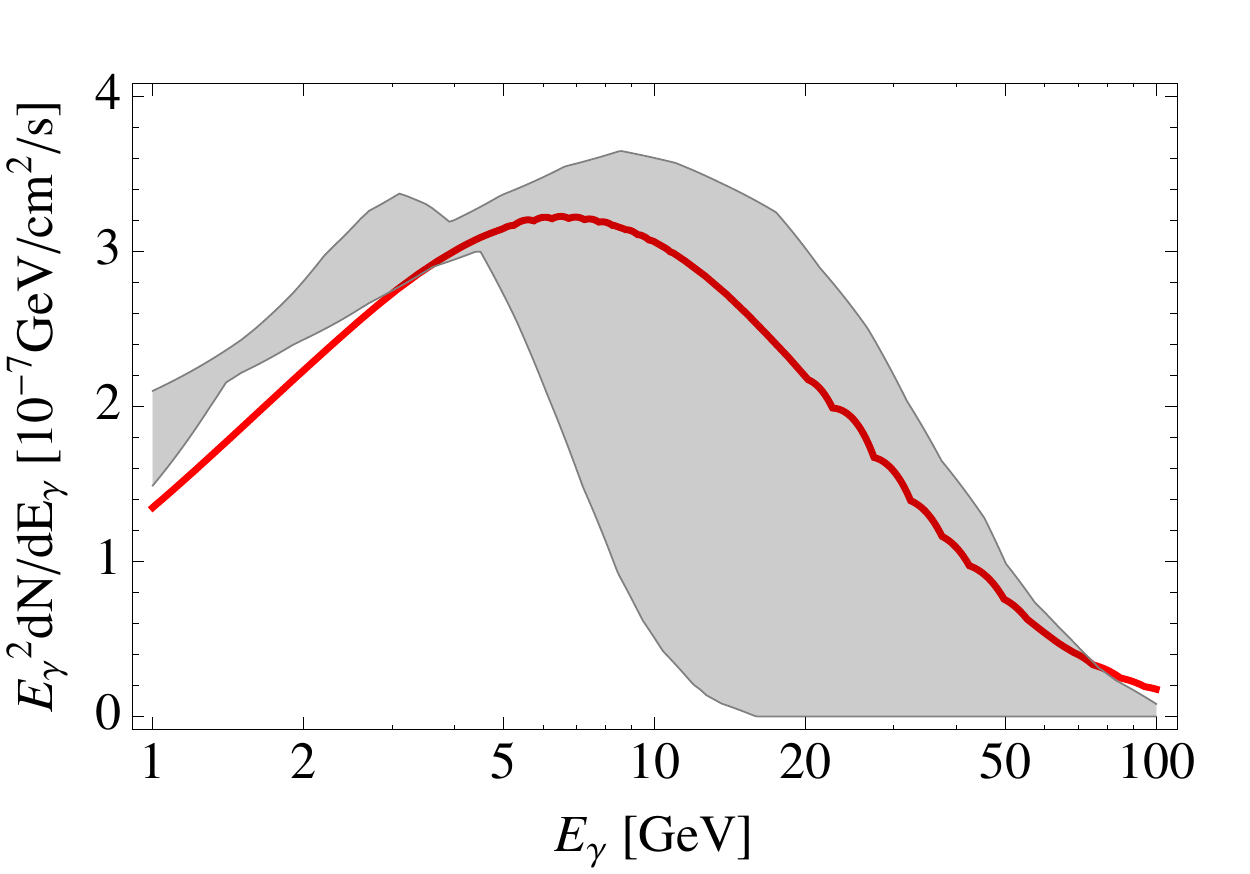} 
   \caption{\emph{Left}: Parameter space of Higgs-portal DM.  The red region is where the annihilation cross section is with 10\% of $2.2\times 10^{-26}$ cm$^3$ s$^{-1}$.  The contours denote the total annihilation cross section of DM into a combination of $WW$, $ZZ$, $hh$ and $t\bar{t}$ final states.  The shaded region is excluded by LUX assuming $\chi$ makes up all the dark matter, $f_\chi=1$. The spectrum for the marked point in the parameter space is shown on the right. \emph{Right}: spectrum for the candidate point.  The annihilation branching fractions are 48\%, 22\%, 28\%, and 2\% for $WW$, $ZZ$, $hh$, and $t\bar{t}$ respectively, and $\mathcal{J}=3$. The grey region represents the envelope of the four Fermi spectra shown in Figure~\ref{fig:at-rest}.}
   \label{fig:pospelonplots}
\end{figure}

In the left panel of Figure~\ref{fig:pospelonplots} we show the parameter space for this model. 
The annihilation cross section into the various channels, as well as the direct detection cross section, can be found in~\cite{Cline:2013gha}. For direct detection we show the limits assuming the effective Higgs-nucleon coupling from~\cite{Junnarkar:2013ac} which is $f_N=0.29$, with a few percent uncertainties. We see that a thermal relic scalar dark matter is consistent with direct detection limits so long as the dark matter mass is above $\sim$125~GeV. From Figure~\ref{fig:Fermi-money} we see that for DM masses above 150 GeV the required cross section to be consistent with the Fermi best fit spectra is significantly higher than the thermal cross section, even after the uncertainty in the line-of-sight integral $\mathcal{J}$ is considered. We note however, that Figure~\ref{fig:Fermi-money} is not  a proper fit to Fermi data and that once all uncertainties are considered lower cross sections may be allowed. Even if the Fermi region remains at higher cross section the GCE can easily be consistent with Higgs portal DM if the couplings $\lambda$ is taken to be larger than its thermal value. In this case the thermal abundance of $\chi$ is lower, but the full DM abundance may come from a non-thermal source, such as a late decay of another thermal relic. Such a scenario is consistent with limits from direct detection, if $f_N$ is sufficiently small. This is seen in Figure~\ref{fig:pospelonplots}, where the right panel shows the spectrum produced by DM annihilations at the parameter point marked as a blue dot on the left for $\mathcal{J}=3$. 

It is interesting to ask - is it possible to get dark matter annihilating dominantly to Higgs bosons? In the framework of the minimal Higgs portal model the annihilation into gauge bosons scales as $\lambda^2$. For large values of $\lambda$ the annihilation into Higgs bosons scales as $\lambda^4$ because of a diagram in which $\chi$ is exchanged in the $t$-channel.
We find that for $m_\chi=170$~GeV the Higgs branching fraction goes above 50\% for $\lambda\sim1.2$ and climbs quickly for larger values of the coupling. As is clear from figure~\ref{fig:pospelonplots}, such large values of $\lambda$ are in conflict with direct detection limits. 

It is possible to get a dominant annihilation into Higgs pairs while evading LUX limits by setting the $h\chi\chi$ coupling to be small while keeping a large $hh\chi\chi$ coupling. Because the Higgs field gets a vev, this can be achieved via tuning, for example, between a quartic interaction and a higher dimension operator. 
Consider, for example, the following couplings to the Higgs portal, including dimension 6 terms:
\begin{equation}
-{\cal L} \supset  \frac{1}{2}\lambda' \, \chi  \, \chi \, \left( H^\dag H - \frac{v^2}{2}\right) 
+ \frac{1}{2\Lambda^2} \, \chi \, \chi \, \left( H^\dag H - \frac{v^2}{2}\right)^2 + \dots
\end{equation}
where $\Lambda$ is the new physics scale at which the effective operator is generated. Notice that while the renormalizable coupling $\lambda'$ leads to a trilinear $h \chi \chi$ coupling, the dimension 6 coupling does not, and instead the leading interaction is the quartic $h h \chi \chi$ coupling. Thus, if one assumes 
$\lambda'$ is small in comparison to $v^2/\Lambda^2$, then the dominant annihilation channel will be $\chi \chi \rightarrow hh$ and limits from direct detection experiments are evaded. We emphasize that the small $\lambda'$ regime indeed corresponds to a tuning of the parameters in the effective theory, as it requires a coincidental relation between the $ \chi^2 |H|^2$ and $\chi^2 |H|^4$ operators in the Lagrangian.  Still, it would be interesting to investigate in more detail scenarios of this kind in relation to a possible explanation of the gamma-ray excess.

\subsubsection{The $h\to\gamma\gamma$ Line}\label{sec:line}
Since we have discussed models in which DM annihilates into Higgses, it is worthwhile to pause to discuss the line feature from $h\to\gamma\gamma$ at 62.5 GeV. 
Because the line is very sensitive to small boosts of the Higgs, a photon line is only present in a sliver of parameter space, at best. In order for a line-search to be effective, the feature being searched for cannot be broader than the energy resolution of the instrument. We point out that the dark matter mass would need to be less than 0.1\% above the mass of the Higgs in order for the boost not to broaden the line width beyond 10\%, the energy resolution of the Fermi instrument around 63 GeV. 
However the signal it produces is sufficiently interesting to consider its observability in the hopes that a mechanism for production of Higgs at rest would be found.

Observation of such a line at $\sim62.5$ GeV would be a spectacular smoking gun signal of DM annihilation to a Higgs. One could have worried that the branching fraction of Higgs to 2 photons is too small to be observed. However the photons produced in this decay are harder and the cosmic ray backgrounds in this region are small. In figure~\ref{fig:at-rest} this is accounted for by the factor of $E^2$ which accentuates the line. 

The recent line search by the Fermi collaboration~\cite{Ackermann:2013uma} places constraints on such spectral features. The search is optimized for a variety of dark matter profiles (following~\cite{Weniger:2012tx}). Among these is a contracted NFW profile (dubbed NFWc) which corresponds to our equation~(\ref{eq:NFW}) with a slope of $\gamma=1.3$ and with a local dark matter density of 0.4 GeV/cm$^3$. Fortunately this profile is compatible with that which is suggested by the GC excess. The bound for a line at 62 GeV is expressed as a limit on the annihilation cross section of dark matter with this mass annihilating to photons. The NFWc Fermi limit is $\langle \sigma v\rangle^{\chi\chi\to\gamma\gamma}  < 3\times 10^{-29}~\mathrm{cm^2/sec}$ for a line at half the Higgs mass. 

Assuming DM annihilates only to $hh$ at rest we can translate the rate to 62.5 GeV photons into an effective cross section of $\chi\chi\to\gamma\gamma$ at that mass
\begin{equation}
\langle\sigma v\rangle_\mathrm{eff}^{\chi\chi\to\gamma\gamma} 
= \frac{1}{4}\langle \sigma v \rangle^{\chi\chi\to hh} \times 2\, \mathrm{BR}(h\to\gamma\gamma)
\end{equation}
what the factor of 1/4 accounts for the fact that dark matter at 125 GeV is twice as dilute at DM at half that mass. The factor of two accounts for the presence of two Higgses in the final state. Using $\langle \sigma v \rangle^{\chi\chi\to hh}\sim 5 \times 10^{-26}$ and 
$\mathrm{BR}(h\to\gamma\gamma)=2.3\times 10^{-3}$ we get an effective $\gamma\gamma$ cross section of
\begin{equation}
\langle\sigma v\rangle_\mathrm{eff}^{\chi\chi\to\gamma\gamma} \sim 5\times 10^{-29}~\mathrm{cm^2/sec}
\end{equation}
We see that despite the tiny branching fraction, the Fermi line search is probing the predicted rate for DM going to Higgses at rest. We stress however, that even if one were to tune the mass of the dark matter to be exceedingly close to the Higgs mass, the phase space for this annihilation process would be tiny and the required rate would not be achievable for a perturbative model. We leave the construction of models that would produce an $h\to\gamma\gamma$ line while explaining the GC excess for future work.

\subsection{Other Models}

There are of course many other possibilities for WIMPs annihilating to electroweak bosons and tops besides the well-motivated SUSY and Higgs portal scenarios discussed above. Here we mention a few of these possibilities:  

\begin{itemize}

\item {\bf Singlet-Doublet Dark Matter}

Another simple scenario leading to thermal relic DM candidates that can explain the GCE is the singlet-doublet model~\cite{ArkaniHamed:2005yv,Mahbubani:2005pt,D'Eramo:2007ga,Enberg:2007rp,Cohen:2011ec}, in which scalar or fermionic dark matter is an admixture of an electroweak singlet and a doublet. There are important differences to the MSSM neutralino discussed in section~\ref{sec:neutralino}, which is tightly constrained by SUSY. In particular, in the MSSM the gaugino-Higgs-higgsino couplings are gauge interactions.  This implies, for example, that mass splittings among charged and neutral state cannot be too large and that co-annihilation is often important. As a result a thermal relic neutralino can satisfy the required rate for the GCE in small regions of parameter space.  In contrast, for a generic singlet-doublet scenario, we are free to take the singlet and doublet coupling to the Higgs to be large, giving larger mass splittings and reducing the amount of co-annihilation.

\item {\bf SUSY with Large $\tan\beta$}

Though we have focused on annihilations to heavier states, it is interesting to remember that the Fermi uncertainties also expand the mass window in which annihilations to bottom quarks can explain the excess. It is known that if dark matter interactions are mediated by a pseudo-scalar one can account for the Gooperon rate while evading direct detection limits~\cite{Boehm:2014hva,Alves:2014yha,Berlin:2014tja,Ipek:2014gua,Cheung:2014lqa}. If dark matter can indeed be heavier, we may be able to use the MSSM's pseudo-scalar Higgs as the mediator. In this case, with large $\tan\beta$, dark matter would annihilate into bottom quarks. It is interesting to explore whether one can build such a model while satisfying LHC bounds from $A\to\tau\tau$ searches, limits on charged Higgses, and direct detection limits coming from the exchange of the heavy scalar Higgs.

\item {\bf Flavored Dark Matter}

A model which is qualitatively similar to the bino-stop model considered earlier is top-flavored dark matter~\cite{Batell:2011tc,Agrawal:2011ze,Batell:2013zwa,Agrawal:2014una,Agrawal:2014aoa,Kile:2011mn,Kamenik:2011nb,Kumar:2013hfa,Gomez:2014lva,Bishara:2014gwa,Hamze:2014wca}. In this framework DM is the lightest component of a flavor multiplet, and under the assumption of minimal flavor violation~\cite{D'Ambrosio:2002ex}, there is an accidental $Z_3$ flavor triality symmetry that is responsible for the DM stability. Bottom-flavored dark matter has previously been investigated as a model of $\chi \bar \chi \rightarrow b \bar b$ which can explain the GCE~\cite{Agrawal:2014una}. 
For the $t\bar t$ final state, top flavored dark matter provides a well-motivated candidate. The basic interaction governing the phenomenology is ${\cal L} \supset \lambda \bar t_R \phi \chi_t +{\rm h.c.}$, which is similar to the bino-stop-top coupling (\ref{eq:bino-stop}) although here $\lambda$ is a free parameter. Because of this freedom, it is possible to obtain larger annihilation cross sections such as those preferred by the Fermi GCE models. Also, since top flavored dark matter is Dirac, there is vector coupling to the $Z$-boson induced at one-loop, leading to a detectable SI scattering cross section at LUX and future ton-scale experiments.

\item  {\bf  Two-Stage Annihilation}

Finally, besides the direct annihilation of dark matter to SM final states which has been the focus of this work, there is the class of models in which dark matter first annihilates to an unstable intermediate state, $\chi \chi \rightarrow \phi \phi$, and then $\phi$ decays to $WW$, $ZZ$, $hh$, or $t\bar t$. This mechanism has already been explored in the literature as an explanation for the GCE with light SM final states such as $b$-quarks or jets~\cite{Boehm:2014bia,Martin:2014sxa,Berlin:2014pya}. This will allow for even heavier dark matter in the few hundred GeV range to account for the GCE. 
However, since the coupling of the intermediate state to the SM can be very small it is difficult to make any firm predictions for direct detection or collider signatures.  

\end{itemize}

\section{Conclusions}\label{sec:conclusion}

Many well-motivated WIMP models feature DM annihilation to weak gauge
bosons, Higgs bosons, and top quarks. In this paper we have explored
possible connections between these final states and a
potential gamma-ray signal. This signal was first observed in the Fermi data
by Goodenough and Hooper~\cite{Goodenough:2009gk,Hooper:2010mq}, and has been confirmed by a number of other
groups -- most recently by the Fermi collaboration. While the presence
of an excess has remained robust, the characterization of the GCE has
steadily evolved in recent years as it has been subjected to
increasingly sophisticated analyses. In particular, recent studies of
the systematic uncertainties in the gamma-ray background models
indicate that GCE gamma-rays may be significantly harder than
previously thought.
In contrast to previous DM interpretations of this signal, which have
focused on $b \bar b$, $\tau \bar \tau$, and light quark final states,
in this paper we have argued that a much broader range final states
and model interpretations, including the annihilation of weak-scale DM
to $WW\,,ZZ\,,hh$, and $t\bar t$ can explain the GCE. 

Our DM interpretations require a quantitative estimate of the
systematic uncertainties in the Fermi GCE, for which we have utilized
two complementary approaches. First, we have used the spectra
determined in the recent study of CCW~\cite{Calore:2014xka}, in which
the background systematics were estimated by a thorough study of a
large range of diffuse emission models and an analysis of the
residuals in several test regions along the Galactic plane. 
Second, we have used the preliminary results of the Fermi analysis of
the Galactic center gamma-ray emission~\cite{simonatalk}. This
analysis employed four distinct diffuse emission models, finding in
all cases that additional NFW templates with cutoff power law spectra
significantly improve the agreement with data. These four best-fit
spectra differ dramatically at energies above 2 GeV, providing a
second independent estimate of the systematic uncertainties in the
GCE. 

With these spectra in hand, we have explored several well-motivated
WIMP models predicting annihilation to $WW\,,ZZ\,,hh$, and $t \bar t$,
including the MSSM neutralino and Higgs portal DM. We have identified
parameters where thermal or non-thermal DM making up all or a
subdominant component could describe the excess. In particular, for
thermal WIMPs saturating the DM abundance, we have found that within
the MSSM a mixed neutralino annihilating to $W$ bosons, or a bino
annihilating to tops via light stop exchange, can explain the GCE.
Interestingly, a light, nonthermal (almost-pure) wino or higgsino, that
forms  a subdominant fraction of the cosmic DM, can also explain the
GCE due to its large annihilation cross section to weak gauge bosons. 
While we have explored only a few slices of the MSSM, it would be
very interesting to investigate in more detail the parameter space
that can describe the GCE. The WIMP models we have investigated in
this work also lead to rich and complementary signatures in direct
detection and collider experiments.  Furthermore, these new final
states predict associated indirect detection signals, such as
antiprotons and (non-GCE) gamma-rays such as those emanating from
dwarf galaxies, whose spectra are different from those expected for
$b\bar{b}$ or $\tau\tau$. We have only scratched the surface of this
exciting interplay, and much remains to be done to correlate these
signatures with the GCE explanation. 

The Galactic center is of course a complex region and much work remains to better characterize the various sources of gamma-rays. With these developments, we expect further refinement of the morphological and spectral properties of the GCE, which in turn will affect any potential DM interpretation. At the moment however, if we take seriously a dark matter interpretation of the excess, then it is too early to conclude whether this signal comes from light DM in the 10-50 GeV range or heavier weak scale DM. We encourage the Fermi collaboration to test all well-motivated interpretations of this signal, including dark matter annihilating to SM final states such light quarks and leptons, top quarks, weak gauge bosons, and Higgs bosons.  

\subsection*{Acknowledgments}

We thank Marco Cirelli, Antonio Delgado, Rouven Essig, Ramona Gr$\ddot{\rm o}$ber, Jack Kearney, Dan Hooper, Travis Martin, Sunghoon Jung, Tracy Slatyer, and Tim Tait for useful discussions and correspondence. We especially thank Ilias Cholis and Christoph Weniger for numerous discussions and information about their recent work~\cite{Calore:2014xka}, and Simona Murgia for discussions about the Fermi analysis~\cite{simonatalk}. 
Fermilab is operated by Fermi Research Alliance,
LLC under Contract No. DE-AC02-07CH11359 with the United States Department of Energy. B.B. is supported in part by a CERN COFUND fellowship. This work was supported in part by the National Science Foundation under Grant No. PHYS-1066293 and the hospitality of the Aspen Center for Physics.

\bibliographystyle{JHEP}
\bibliography{gooperon}

\end{document}